\documentclass{IEEEtran}
\usepackage{cite}
\usepackage{amsmath,amssymb,amsfonts}
\usepackage{algorithmic}
\usepackage{graphicx}
\usepackage{textcomp}
\usepackage{subfigure}
\usepackage{arydshln}
\usepackage{color}
\def\BibTeX{{\rm B\kern-.05em{\sc i\kern-.025em b}\kern-.08em
    T\kern-.1667em\lower.7ex\hbox{E}\kern-.125emX}}

\newtheorem{theorem}{Theorem}[section]
\newtheorem{lemma}[theorem]{Lemma}

\newtheorem{defn}[theorem]{Definition}
\newtheorem{remark}[theorem]{Remark}

\begin{document}
\title{Distributed control of multi-consensus}
\author{Lucia Valentina Gambuzza, Mattia Frasca, \IEEEmembership{Senior, IEEE}
\thanks{L. V. Gambuzza and M. Frasca are with the Dipartimento di Ingegneria Elettrica Elettronica e Informatica, University of Catania, Italy. E-mail: [lucia.gambuzza,mattia.frasca]@dieei.unict.it.}
}


\maketitle

\begin{abstract}
We consider the problem of steering a multi-agent system to multi-consensus, namely a regime where groups of agents agree on a given value which may be different from group to group. We first address the problem by using distributed proportional controllers that implement additional links in the network modeling the communication protocol among agents and introduce a procedure for the optimal selection of them. Both the cases of single integrators and of second-order dynamics are taken into account and the stability for the multi-consensus state is studied, ultimately providing conditions for the gain of the controllers. We then extend the approach to controllers that either add or remove links in the original structure, by preserving eventually the weak connectedness of the resulting graph.
\end{abstract}


\section{Introduction}
\label{sec:introduction}

Since the seminal papers \cite{jadbabaie2003coordination,olfati2004consensus,ren2005consensus}, the consensus problem has received great attention from various perspectives (consensus in linear and nonlinear multi-agents systems, finite-time consensus, stochastic consensus) and in different fields (control engineering, physics, opinion dynamics, biology, among the others). In a system of interacting agents consensus corresponds to the condition in which all the agents converge to a common value. In applications such as robot formation control, flocking, rendez-vous problems, decision making \cite{fax2004information,liu2003stability,bullo2009distributed}, consensus represents the target of the control as it indicates that the units of the system are operating in a coordinated way; for this reason, strategies for designing the communication protocol such that the consensus state is stable have been widely investigated \cite{ren2010distributed,mesbahi2010graph}.

However, there are other applications, as well as specific instances of those previously mentioned, requiring that the behavior of the units is differentiated into small subgroups. For example, formation of a team of robots may need to be split into smaller subformations to simultaneously accomplish several tasks or the temperature of a building to be controlled such that the rooms of different floors have distinct set points \cite{andreasson2014distributed}. This scenario is referred to as multi-consensus or cluster consensus and is characterized by parts of the multi-agent system simultaneously reaching different consensus states \cite{yu2010group}. The importance of multi-consensus is not limited to engineering applications; it is, for instance, momentous in brain science where, thanks to the connectivity and the structure of the brain, each area could perform specific task \cite{schnitzler2005normal}, as well in other natural systems, e.g., bird flocks or schools of fish splitting into different subgroups for avoiding predation or for foraging. Examples of multi-consensus are also found in social systems, e.g., the dynamics of different coexisting opinions or pattern formation in bacteria colonies \cite{blondel2010continuous,you2009collective}.

\subsection{Literature review}

Previous works on the analysis of multi-consensus have investigated the properties of the network of interaction among agents leading to multi-consensus states \cite{chen2011cluster,klickstein2019symmetry,monaco2019multi,xia2011clustering}. More in details, the criteria found in \cite{chen2011cluster} are based on the use of Markov chains and nonnegative matrix analysis for fixed and switching topologies, while the existence of multi-consensus is related to the presence of symmetries \cite{klickstein2019symmetry} or of external equitable partitions \cite{monaco2019multi} in the topology. Finally, multi-consensus can be observed also in the presence of delays or differentiation of the dynamics of the units, as shown in \cite{xia2011clustering}.

Previous results on the control of multi-consensus have considered the control of clusters already existing in the network structure \cite{qin2013cluster} or the introduction of different external inputs to differentiate the dynamics of the clusters \cite{han2013cluster}. In our paper, instead, we deal with the open problem of modifying the structure of the network of interaction in a multi-agent systems such that to obtain arbitrarily selected clusters.

A problem related to multi-consensus is that of cluster synchronization, where the dynamics of the units is oscillatory and thus nonlinear. Similarly to what occurs for multi-consensus, also for cluster synchronization the
appearance of groups of nodes converging to the same behavior has been linked to the existence of symmetries
\cite{pecora2013symmetries,sorrentino2016complete} or external equitable partitions \cite{schaub2016graph,gambuzza2019criterion} in the network structure. These approaches rely on computational group theory or graph theoretical methods, while contraction theory has been applied to derive sufficient conditions for cluster synchronization in \cite{aminzare2018cluster}. A different method exploits the symmetries in the node dynamics rather than in the topology to induce some desired pattern of synchronization \cite{fiore2017exploiting}. The stability of cluster synchronization in weighted networks of heterogeneous Kuramoto oscillators is, instead, studied in \cite{menara2019stability}.

Finally, it is worth to mention that the problem dealt with in this paper is also connected to affine formation control, which targets at stabilizing a collection of states for a multi-agent system that can be associated with a target configuration through an affine transformation and that has been recently solved through local interactions in \cite{lin2015necessary}.

\subsection{Statement of contribution}

Aim of this work is to introduce a strategy for designing distributed proportional controllers to achieve an arbitrary multi-consensus in a multi-agent system. Our technique relies on the notion of external equitable partitions (EEPs) to modify the structure of interactions among agents and can be applied either i) by designing (or re-designing) offline the network of interactions of the multi-agent system and then implementing the needed changes performed by the controllers or ii) by implementing a cyber-layer of controllers operating in parallel with the physical connections of the multi-agent system and providing the further inputs generated by the coupling terms of the control layer. In particular, we first address the problem by considering only the addition of links to the original structure and investigating both the cases of single integrator and second-order dynamics. Here, for single-integrator dynamics we leverage the results of \cite{monaco2019multi} regarding the analysis of the multi-consensus state in networks with EEPs and propose a solution for the control problem. Instead, in the case of second-order dynamics, we also perform the analysis of the stability of the multi-consensus state, ultimately providing a conditions on the values of the gains used in the communication protocol complementing the topological condition linked to the existence of an EEP with given properties. Finally, we extend our method to the case where links may be either added or removed. Our main contributions can be summarized as follows:

\begin{enumerate}
\item introduction of a technique to modify the structure of a network by the addition of new links such that an arbitrary EEP forms (Lemma \ref{lem:lemma1});

\item solution of the multi-consensus control problem through distributed proportional controllers for the case of single integrators (Theorem \ref{th:th1});

\item solution of the multi-consensus control problem through distributed proportional controllers for the case of second-order dynamics (Theorem \ref{th:mainforthesecondorder});

\item introduction of a technique to modify the structure of a network by the addition/removal of links such that an arbitrary EEP forms (Lemmas \ref{lem:lemma1signed} and \ref{lem:lemma1signedconnected}).

\end{enumerate}

The rest of the paper is organized as follows: in Sec. \ref{sec:preliminaries} some preliminary notions are given; in Sec. \ref{sec:problemformulation} the multi-consensus control problem is formulated; in Sec. \ref{sec:singleint} the solution for multi-agent systems of single integrators is discussed, while in Sec. \ref{sec:secondOrder} the case of multi-agent systems with second-order dynamics is dealt with; in Sec. \ref{sec:signedLaplacians} the extension to the case of addition/removal of links is illustrated; in Sec. \ref{sec:conclusions} the conclusions of the paper are drawn.

\section{Preliminaries}
\label{sec:preliminaries}

In this section, we recall some definitions and results of matrix analysis \cite{horn1994matrix}, graph theory \cite{estrada2012structure,latora2017}, group symmetry \cite{heine2007group}, and equitable partitions \cite{godsil2013algebraic}, which will be used throughout the paper.

\subsection{Matrix analysis}

We indicate by $\mathrm{I}_n$ the identity matrix with dimension $n \times n$; with $\mathrm{0}_{m \times n}$ a matrix of zeros of dimension $m \times n$; with $\mathbf{1}_n$ the vector of dimension $n$ with unitary elements. A diagonal matrix, say $\mathrm{D}$, with diagonal terms $\lambda_1, \ldots, \lambda_n$ is indicated as $\mathrm{D}=\mathrm{diag} \{\lambda_1, \ldots, \lambda_n\}$. Moreover, for a positive semidefinite matrix $\mathrm{A}$ we indicate as $\lambda_2(\mathrm{A})$ its smallest nonzero eigenvalue. Given a rectangular matrix $\mathrm{A}$, $\mathrm{A}^+$ indicates its Moore-Penrose pseudoinverse.

We recall a property on the eigenvalues of the product of two matrices that is used in the derivation of our results.

\begin{lemma}{(Theorem 1.3.22 in \cite{horn2012matrix})}\label{lemma:sameEigenv}
Let $\mathrm{A} \in \mathbb{R}^{m \times n}$ and $\mathrm{B} \in \mathbb{R}^{n \times m}$ with $m\leq n$. Then the $n$ eigenvalues of $\mathrm{B} \mathrm{A}$ are the $m$ eigenvalues of $\mathrm{A} \mathrm{B}$ together with $n-m$ zeros. In particular if $\mathrm{A}$ and $\mathrm{B}$ have same dimension, $m=n$, then $\mathrm{B} \mathrm{A}$ has the same eigenvalues of $\mathrm{A} \mathrm{B}$.
\end{lemma}

We also introduce the definition of vectorization used for the derivation of the optimization problem.

\begin{defn}{(\cite{macedo2013typing})}
Vectorization is a linear transformation which converts a matrix $\mathrm{A}$ into a column vector $vec(\mathrm{A})$, which corresponds to parsing $\mathrm{A}$ in column-major order, e.g.,

\[
 \mathrm{A} =\left[
  \begin{array}{ccc}
   a_{11} & a_{12} & a_{13} \\
   a_{21} & a_{22} & a_{23}
  \end{array}\right]
  \Rightarrow
  vec(\mathrm{A}) =\left[
  \begin{array}{c}
   a_{11} \\ a_{21} \\ a_{12} \\
   a_{22} \\ a_{13} \\ a_{23}
  \end{array}\right]
\]
\end{defn}

\subsection{Graph theory}

\begin{defn}[directed graph/digraph]
A directed graph $\mathcal{G}$, shortly a digraph, is defined by the set of nodes/vertices $\mathcal{V(G)}=\{v_1,\ldots,v_N\}$, and the set of directed edges/links $\mathcal{E(G)} \subseteq \mathcal{V}$ x $\mathcal{V}$. A directed edge from node $v_i$ to node $v_j$ is represented as an ordered pair $(v_i, v_j)$, indicating that agent $v_j$ can obtain information from agent $v_i$.
\end{defn}



To indicate the nodes of $\mathcal{G}$ we will equivalently use $v_i$ or, shortly, $i$.
We said that node $i$ is neighbor of node $j$ if there exist the arc from $i$ to $j$, we also denote by $\mathcal{N}_i=\{v_j \in \mathcal{V} : (v_i \rightarrow v_j) \in \mathcal{E}\}$ the set of neighbors of node $i$. We indicate the cardinality of a set $\mathcal{N}$, i.e., the number of elements contained in it, as $|\mathcal{N}|$.


The digraph $\mathcal{G}$ can be fully represented by its adjacency and Laplacian matrices. The elements of the adjacency matrix $\mathrm{A}$ are defined as:

\[
 a_{ij} =
  \begin{cases}
   1 & \text{if } (v_j,v_i) \in \mathcal{E} \\
   0       & \text{otherwise}
  \end{cases}
\]
We assume that there are no self-loops, i.e., $a_{ii}= 0$ for all $i=1,\ldots,N$. We define the degree $k_i$ of node $i$ as the number of connections incident on node $i$: $\sum_{\substack{j=1, j\neq i}}^N {a}_{ij} = k_i$, with $i=1,2,\ldots,N$.
Correspondingly we define the in-degree Laplacian matrix $\mathrm{L}({\mathcal{G}})$, whose elements are ${l}_{ij}=-a_{ij}$ if $i \neq j$, and ${l}_{ii}= k_i$. We denote with $\lambda(\mathrm{L})$ the set of eigenvalues of $\mathrm{L}$.

A directed path of length $m$ in $\mathcal{G}$ is given by a sequence of distinct vertices $v_{i0}, v_{i1}, \ldots, v_{im}$ such that for $k=0,1,\ldots,m-1$ the vertices $(v_{ik},v_{ik+1}) \in \mathcal{E}$ \cite{mesbahi2010graph}. A digraph is called strongly connected if for every pair of vertices there is a directed path between them. The digraph is called weakly connected if it is connected when viewed as a graph, that is, when the disoriented graph is connected. It is called rooted if it is weakly connected and contains at least one rooted out-branching.

\subsection{Equitable partitions}

Equitable partitions represent regularities in the structure of the underlying topology of interactions between agents that are reflected into the dynamical collective state that the network generates, an issue that has been widely investigated in the context of cluster synchronization, both in terms of analysis \cite{schaub2016graph} and control \cite{gambuzza2019criterion}. The concept of equitable partitions has been found fundamental also for the study of minimum time for convergence \cite{yuan2013decentralised}, controllability \cite{martini2010controllability} and observability \cite{o2013observability} in consensus of multi-agent systems.

Given a graph $\mathcal{G}$ and the set of vertices associated to it $\mathcal{V}=\mathcal{V}(\mathcal{G})$, a partition $\pi$ is a map of the vertices that groups them into $M$ distinct cells, $C_1, C_2, \ldots, C_M$, with $\bigcup \limits_{l=1}^M C_l=\mathcal{V}$ and $C_i \cap C_j = \emptyset$, for $i \neq j$.

\begin{defn}
A partition $\pi=\{ C_1, C_2, \ldots, C_M \}$ is said equitable if, for any pairs of cells $C_l$ and $C_k$ (including $l=k$), there exists a constant $b_{lk}$ such that each vertex $v_i \in C_l$ has exactly $b_{lk}$ neighbors in $C_k$.
\end{defn}

The notion of equitable partition thus requires that nodes inside a cell have the same out-degree pattern with respect to any other cell (including internal links). Given an equitable partition $\pi$, the quotient graph of $\mathcal{G}$ over $\pi$, denoted by $\mathcal{G}/\pi$, is the directed graph with vertices $C_1, C_2, \ldots, C_M$ and $b_{lk}$ arcs from $C_l$ to $C_k$. This graph is regular.

External equitable partitions (also known as almost or relaxed equitable partitions) constitute a relaxed version of equitable partitions, formally expressed given by the following definition.

\begin{defn}
Given a graph $\mathcal{G}$, and a partition $\pi =\{ C_1, C_2, \ldots, C_M \} $ of the vertex set $\mathcal{V(G)}$, if for any pairs of cells $C_l$ and $C_k$, with $l \neq k$, each vertex $v_i \in C_l$ has exactly $b_{lk}$ neighbors in $C_k$, thus $\pi$ is an external equitable partition (EEP).
\end{defn}

In EEPs it is not important that the graph induced by the partition is regular, thus nodes within a cell do not necessarily have the same numbers of neighbors. While the cells of an equitable partition have the same out-degree pattern with respect to every cell, in EEPs this holds only for the number of connections between distinct cells. Each network always has two trivial EEPs: $\pi= \{ \{1\}, \{2\}, \ldots, \{N\} \}$, where all the cells are singletons containing a single node, and $\pi= \{\{1, 2, \ldots, N \} \}$, where all the nodes are grouped into a single cell.

The characteristic matrix $\mathrm{P}$ of an EEP $\pi= \{C_1, C_2, \ldots, C_M \}$, is the $N \times M$ matrix with $p_{ij}=1$ if node $i$ belongs to cell $C_j$, and $p_{ij}=0$ otherwise. The characteristic matrix $\mathrm{P}$ is such that $\mathrm{P}^T\mathrm{P}$ is diagonal with the $j$-th element equal to the number of vertices in the cell $C_j$. Since the diagonal terms are nonzero as the cells are not empty, $\mathrm{P}^T\mathrm{P}$ is invertible.

Let us indicate the Laplacian matrix of the quotient graph as $\mathrm{L}^{\pi}$; we have that $\mathrm{L}\mathrm{P}=\mathrm{P}\mathrm{L}^{\pi}$, and  $\mathrm{L}^{\pi}=\left(\mathrm{P}^T\mathrm{P}\right)^{-1}\mathrm{P}^T\mathrm{L}\mathrm{P}$ \cite{o2013observability}. In addition, if $\pi$ is an EEP of $\mathcal{G}$, then, the eigenvalues of the $\mathrm{L}^{\pi}$ are a subset of the eigenvalues of $\mathrm{L}$, $\lambda(\mathrm{L}^{\pi})\subseteq \lambda(\mathrm{L})$, and $v^{\pi}$ is an eigenvector of $\mathrm{L}^{\pi}$ if $v$ ($v=P v^{\pi}$) is an eigenvector of $\mathrm{L}$ with the same eigenvalue.

From the characteristic matrix $\mathrm{P}$ the projection operator into the cell subspace is defined, i.e., $\mathrm{P}_H=\mathrm{P}\left(\mathrm{P}^T\mathrm{P}\right)^{-1}\mathrm{P}^T$. This operator is linked to the Laplacian $\mathrm{L}$ by the relation

\begin{equation}
\label{eq:PH}
\mathrm{L} \mathrm{P}_H=\mathrm{P}_H \mathrm{L}\mathrm{P}_H
\end{equation}

We now recall the definitions of reaches of a digraph.

\begin{defn}[\cite{caughman2006kernels}]
A reachable set $\mathcal{R}(v_i)$ of a vertex $i$ is the set containing node $v_i$ and all the nodes $v_j$ such that there exists a path from $v_i$ to $v_j$.
\end{defn}

A set $\mathcal{R}$ of vertices is called a reach if $\mathcal{R} = \mathcal{R}(v_i)$ for some $v_i$ and there is no vertex $v_j$ such that $\mathcal{R}(v_i) \not \subseteq \mathcal{R}(v_j)$. For each reach $\mathcal{R}(v_i)$ of a graph, we define the exclusive part of $\mathcal{R}(v_i)$ as the set $\mathcal{H}_i= \mathcal{R}(v_i) \backslash \bigcup \limits_ {v_j\neq v_i} \mathcal{R}(v_j)$. Likewise, we define the common part of $\mathcal{R}(v_i)$ to be the set $\mathcal{C}=\mathcal{R}(v_i) \backslash\ \mathcal{H}_i$.

By definition it follows that the pairwise intersection of two exclusive sets is empty, i.e., $\mathcal{H}_i \cap \mathcal{H}_j=\emptyset$.
For a digraph the number of reaches is equal to the multiplicity $\mu$ of the zero eigenvalue of the Laplacian matrix $\mathrm{L}$.

\section{Problem formulation}
\label{sec:problemformulation}

We study control of multi-consensus for two multi-agent systems (single integrators and second-order dynamics), formulating two different problems.

\subsection{Problem 1. Multi-consensus of single integrators.}

\emph{Model}. We consider a multi-agent system described by:

\begin{equation}
\label{eq:MAS}
\dot{x}_i(t)=-\sum\limits_{j=1}^N\mathrm{L}_{ij}x_j(t)+u_i(t)
\end{equation}

\noindent with $i=1,\ldots,N$. ${x}_i(t)$ represents the state variable of node or agent $i$ at time $t$, $\mathrm{L}_{ij}$ the elements of the Laplacian matrix of the digraph modeling the original connectivity between agents, and $u_i(t)$ distributed proportional controllers:

\begin{equation}
\label{eq:controllers}
u_i(t)=-\sum\limits_{j=1}^N \mathrm{L}^u_{ij}x_j(t)
\end{equation}

\noindent where $\mathrm{L}^u_{ij}$ are the elements of the Laplacian matrix $\mathrm{L}^u$ of a second digraph, that is, the \emph{control layer}, representing the links that are added to the original multi-agent system by the controllers. Defining the stack vectors $\mathbf{x}=[{x}_1,{x}_2,\ldots,{x}_N]^T$ and $\mathbf{u}=[{u}_1,{u}_2,\ldots,{u}_N]^T$, we can equivalently express the control as

\begin{equation}
\label{eq:controlLayer}
\mathbf{u}(t)=-\mathrm{L}^u\mathbf{x}(t)
\end{equation}

\begin{remark}
Equations~(\ref{eq:MAS}) are widely used in multi-agent problems such as flocking, swarming, and distributed estimation involving a scalar variable for each node \cite{mesbahi2010graph}.
\end{remark}

\emph{Problem.} Given a multi-agent system as in (\ref{eq:MAS}) and a partition $\pi^Q=\{C_1, C_2, \ldots, C_Q\}$ of the set of agents into $Q$ cells, $\pi^Q=\{C_1, C_2, \ldots, C_Q\}$ with $C_h \bigcap C_l=0$ for $h,l=1,\ldots,Q$, $h\neq l$ and $\bigcup\limits_{h=1,\ldots,Q} C_h=\mathcal{V}$, find the controllers (\ref{eq:controllers}), i.e., design the \emph{control layer} $\mathrm{L}^u$, such to obtain the multi-consensus defined by:

\begin{equation}
\label{eq:multiconsensus}
\lim\limits_{t\rightarrow +\infty} |x_j(t)-x_i(t)|=0, \forall i,j \in C_h, \forall h=1,\ldots,Q
\end{equation}

\begin{remark}
Note that the multi-consensus is defined by partitioning the set of agents into $Q$ cells and requiring that the state trajectories of agents belonging to the same cell converge asymptotically. Correspondingly, the multi-consensus manifold is defined as $\mathcal{M}=\{\mathbf{x}\in\mathbb{R}^N|x_i=x_j, \forall i,j \in C_h, \forall h=1,\ldots,Q \}$. In this definition we do not impose that different groups have distinct consensus values, but only that units in the same cell converge to the same consensus value. As we will show later, in fact, the proposed approach does not exclude that some cells may merge together.

\end{remark}

\begin{remark}
\label{rem:remark33}
Note also that, if the digraph is strongly connected or weakly connected and rooted, then the associated multi-agent system reaches consensus, which can be viewed as a particular multi-consensus where all cells converge to the same consensus value. In this case, the problem is trivial and admits a solution with $\mathrm{L}^u=0$.
\end{remark}

In the following, unless explicitly stated, we therefore assume that the digraph underlying the multi-agent system is weakly connected and not rooted.

\begin{remark}
As recently pointed out in \cite{panteley2017synchronization}, in the case of consensus in networks of either homogeneous or heterogeneous units, the multi-agent system is governed by an emergent dynamics corresponding
to that of the mean-field unit restricted to the synchronization manifold. In this context, consensus is reached when the motions of all the units converge to that of the emergent dynamics, and is, therefore, studied as a problem of stability of this emergent dynamics, generating a dichotomy between the motion on the consensus manifold and the consensus error. In the case, here discussed, of multi-consensus, the multi-agent system is instead governed by distinct emergent dynamics, corresponding to the motion generated in the cells of the partition.
\end{remark}

\subsection{Problem 2. Multi-consensus of second-order dynamics.}

\emph{Model}. We consider a multi-agent system with second-order consensus dynamics defined as follows. Let $x_i(t), v_i(t) \in \mathbb{R}$ be the state variables of each agent $i$, with $i=1,\ldots,N$ and let $a$ and $b$ be two constant real parameters, then

\begin{equation}
\label{eq:IIorder}
\begin{array}{l}
\dot{x}_i(t)=v_i(t)\\
\dot{v}_i(t)= a x_i(t) + b v_i(t) - \sum\limits_{j=1}^N\mathrm{L}_{ij} (k_1 x_j(t)+k_2 v_j(t)) + u_i(t)
\end{array}
\end{equation}

\noindent where $k_1$ and $k_2$ are constant parameters, referred to as gains of the communication protocol, and $u_i(t)$ is the control input for agent $i$. The agents initially interact each other according to the topology defined by the Laplacian $\mathrm{L}$.

Introducing $\mathrm{K}=[k_1 ~ k_2]$, system (\ref{eq:IIorder}) can be expressed in matrix form as

\begin{equation}
\label{eq:IIorderMatrixForm}
\dot{\mathbf{x}}_i(t)= \mathrm{A} \mathbf{x}_i(t) - \sum\limits_{j=1}^N\mathrm{L}_{ij} \mathrm{B}\mathrm{K}\mathbf{x}_j(t) + \mathrm{B} u_i(t)
\end{equation}

\noindent where $\mathbf{x}_i=\left[ \begin{array}{ll} x_i \\v_i \end{array}\right]$, $\mathrm{A}=\left[ \begin{array}{ll} 0 &1\\a &b \end{array}\right]$, and $\mathrm{B}=\left[ \begin{array}{ll} 0 \\1 \end{array}\right]$. Analogously to problem 1, we consider distributed proportional controllers ${u}_i(t)$

\begin{equation}
\label{eq:controllersIIorder}
{u}_i(t)= -\mathrm{K} \sum \limits_{j=1}^N \mathrm{L}^u_{ij} \mathbf{x}_j(t)
\end{equation}

\noindent such that system (\ref{eq:IIorder}) with the inclusion of the controllers (\ref{eq:controllersIIorder}) becomes

\begin{equation}
\label{eq:II}
\dot{\mathbf{x}}_i(t)= \mathrm{A} \mathbf{x}_i(t) - \sum\limits_{j=1}^N(\mathrm{L}_{ij}+\mathrm{L}^u_{ij} ) \mathrm{B}\mathrm{K}\mathbf{x}_j(t)
\end{equation}

Indicating with $\mathbf{X}=[\mathbf{x}_1^T,\mathbf{x}_2^T,\ldots,\mathbf{x}_N^T]^T$ the stack vector of the state variables of the agents, the control terms included in (\ref{eq:IIorderMatrixForm}) can be equivalently rewritten in terms of the control layer $\mathrm{L}^u$ as:

\begin{equation}
\label{eq:controlLayer}
\mathrm{B}\mathbf{u}(t)=-\mathrm{L}^u\otimes \mathrm{BK} \mathbf{X}(t)
\end{equation}

Note that the controller gains $k_1$ and $k_2$ are assumed to be equal to the gains of the communication protocol and for this reason they are indicated with the same symbol.

\begin{remark} In the context of vehicle dynamics the variables $x_i(t)$ and $v_i(t)$ represent the position and the velocity of agent $i$, and the parameters $a$ and $b$ are the stiffness and damping factor. System (\ref{eq:IIorder}) may be also interpreted as a set of generic second-order linear dynamical units, each in controllable canonical form. For $a=b=0$ it reduces to a multi-agent system of double integrators, modeling for instance the dynamical interactions of space satellites \cite{andreasson2014distributed}.
\end{remark}

\emph{Problem}. Given a multi-agent system as in (\ref{eq:IIorder}) and a partition $\pi^Q=\{C_1, C_2, \ldots, C_Q\}$ of the set of agents into $Q$ cells, find the controllers (\ref{eq:controllersIIorder}), or equivalently design the \emph{control layer} $\mathrm{L}^u$ and the gains $k_1$ and $k_2$, such to obtain the multi-consensus

\begin{equation}
\label{eq:multiconsensusII}
\lim\limits_{t\rightarrow +\infty} \|\mathbf{x}_j(t)-\mathbf{x}_i(t)\|=0, \forall i,j \in C_h, \forall h=1,\ldots,Q
\end{equation}

\begin{remark}
The multi-consensus manifold for the multi-consensus problem of multi-agent systems with second-order dynamics is defined as $\mathcal{M}=\{\mathbf{X}\in\mathbb{R}^{2N}|\mathbf{x}_i=\mathbf{x}_j, \forall i,j \in C_h, \forall h=1,\ldots,Q \}$. The given problem is equivalent to find the controllers (\ref{eq:controllersIIorder}), or equivalently design the \emph{control layer} $\mathrm{L}^u$ and the gains $k_1$ and $k_2$ such that the multi-consensus manifold exists and is stable.
\end{remark}

\section{Control of multi-consensus of single integrators}
\label{sec:singleint}

\subsection{Controller design}

To illustrate the design of the controllers for multi-consensus of single integrators, we first introduce the following lemma, showing how to modify the structure of a network so that it has a given EEP.

\begin{lemma}
\label{lem:lemma1}
Given a network with Laplacian matrix $\mathrm{L}$ and a partition $\pi^Q$, there exists a Laplacian matrix $\mathrm{L}^u$ such that $\pi^Q$ is an EEP for the network with Laplacian matrix $\mathrm{L}+\mathrm{L}^u$.
\end{lemma}

\emph{Proof:} First the characteristic matrix $\mathrm{P}$ of the partition $\pi^Q$  is built. Its elements are fixed as: $\mathrm{P}_{ih}=1$ if agent $i$ belongs to cell $C_h$ and $\mathrm{P}_{ih}=0$, otherwise. From $\mathrm{P}$, the operator $\mathrm{P}_H$ is derived as $\mathrm{P}_H=\mathrm{P}\mathrm{P}^+$.

The key property to find $\mathrm{L}^u$ is expressed by Eq.~(\ref{eq:PH}). Hence, $\pi^Q$ is an EEP for the network with Laplacian matrix $\mathrm{L}+\mathrm{L}^u$ if

\begin{equation}
\label{eq:permute}
(\mathrm{L}+\mathrm{L}^u)\mathrm{P}_H=\mathrm{P}_H(\mathrm{L}+\mathrm{L}^u)\mathrm{P}_H
\end{equation}

This yields that

\begin{equation}
\label{eq:permute2}
\mathrm{L}^u\mathrm{P}_H-\mathrm{P}_H \mathrm{L}^u\mathrm{P}_H = \mathrm{P}_H \mathrm{L}\mathrm{P}_H - \mathrm{L}\mathrm{P}_H
\end{equation}

Equation~(\ref{eq:permute2}) is a Lyapunov equation with unknown $\mathrm{L}^u$. By vectorization it can be recast as

\begin{equation}
\label{eq:permute2vec}
(\mathrm{P}_H^T \otimes  \mathrm{I}_N)vec(\mathrm{L}^u)- (\mathrm{P}_H^T \otimes  \mathrm{P}_H)vec(\mathrm{L}^u) = vec(\mathrm{P}_H \mathrm{L}\mathrm{P}_H - \mathrm{L}\mathrm{P}_H)
\end{equation}

%

Equation~(\ref{eq:permute2vec}) becomes

\begin{equation}
\label{eq:MxB}
\mathrm{M}y=\mathrm{B}
\end{equation}

\noindent with $\mathrm{M}=\left [ (\mathrm{P}_H^T \otimes  \mathrm{I}_N) -  \mathrm{P}_H^T \otimes  \mathrm{P}_H \right ]$, $\mathrm{B}= -\left [  vec({\mathrm{P}_H\mathrm{L}\mathrm{P}_H-\mathrm{L}\mathrm{P}_H})  \right ]$ and $y=-vec(\mathrm{L}^u)$. The vector $y$ comprises elements $y_h$ associated to terms $-\mathrm{L}^u_{ij}$ with $i\neq j$ that may take binary values, i.e., $y_h=\{0,1\}$ for $h=(i-1)N+j$ with $i,j=1,\ldots,N$ and $i\neq j$, while the other elements, which are associated to terms $-\mathrm{L}^u_{ii}$, are constrained by the zero-row sum condition of the Laplacian, i.e., $y_h=-\sum\limits_{l=(i-1)N+1,l\neq h}^{iN} y_l$ for $h=i(N+1)-N$ with $i=1,\ldots,N$.

The existence of a solution for Eq.~(\ref{eq:MxB}) is guaranteed by the following argument. Consider the complete graph $\mathcal{K}$. Its Laplacian matrix is given by $\mathrm{L}_\mathcal{K}=(N-1)\mathrm{I}-\mathbf{1}\mathbf{1}^T$. Replacing $\mathrm{L}+\mathrm{L}^u$ with $\mathrm{L}_\mathcal{K}$ in Eq.~(\ref{eq:permute}) we get a trivial identity, so $\pi^Q$ is an admissible EEP for the complete graph $\mathcal{K}$. Hence, designing a control layer that adds the links to complete the graph always leads to a solution to the problem. However, clearly this is not an efficient solution as likely involves a large number of links. Instead, we look for a solution that, on the contrary, minimizes the number of links of the control layer $\mathrm{L}^u$.

These considerations prompt the definition of an optimization problem with binary variables $y_h$ with $h=(i-1)N+j$, $i,j=1,\ldots,N$, $i\neq j$.

\begin{equation}
\label{eq:LyapEqVecV2minp}
\min f^T y, \mathrm{~subject~to~}\mathrm{M} y=\mathrm{B}
\end{equation}

\noindent where $f_h=1$ for $h=(i-1)N+j$, $i,j=1,\ldots,N$, $i\neq j$ and $f_h=0$, otherwise. The solution of the optimization problem defined above is a Laplacian matrix $\mathrm{L}+\mathrm{L}^u$ satisfying (\ref{eq:permute}). $\square$

\begin{remark}
\label{rem:constructivealgo}
The optimization problem (\ref{eq:LyapEqVecV2minp}) can be solved by using standard integer linear programming solvers \cite{hillier1995introduction} or the following constructive algorithm. First, assign the nodes of the multi-agent system to the $Q$ cells defined by the partition $\pi^Q$. Then, for each pair of cells $C_h$ and $C_k$ with $h,k=1,\ldots,Q$, $h\neq k$, define with $b_{kh}$ the maximum number of links that start from a node in $C_h$ and end in a node in $C_k$, i.e.,  $b_{kh}=\max\limits_i (\sum\limits_i a_{ji}|i\in C_h, j\in C_k)$. Now, taking into account that $\sum\limits_{i\in C_h}a_{ji}$ gives the total number of links that start from $C_h$ and end in node $j \in C_k$, add a number of links equal to $b_{kh}-\sum\limits_{i\in C_h}a_{ji}$ from nodes of $C_h$ to $j$. These nodes need to be not already connected to $j$, i.e., the links should represent new connections, not already existing in the original network (this can be easily checked by inspecting the adjacency matrix of the original network). These considerations also allow to calculate the minimum number of links to add, here indicated as $n_l$. Taking into account that for each pair of cells the algorithm adds a number of links equal to $b_{kh}-\sum\limits_{i\in C_h}a_{ji}$, then $n_l$ is given by: \begin{equation}
n_l=\sum\limits_{k,h=1,k\neq h}^Q(b_{kh}-\sum\limits_{i\in C_h}a_{ji})
\end{equation}
\end{remark}

Lemma~\ref{lem:lemma1} shows that with the addition of new links, formalized through the matrix $\mathrm{L}^u$, it is possible to change the original topology of the multi-agent system such that $\pi^Q$ is an EEP for the new network. We now show that this guarantees to reach the associated multi-consensus. To this aim, leveraging recent results discussed in \cite{monaco2019multi}, we introduce a new partition for the network with Laplacian $\mathrm{L}+\mathrm{L}^u$ and recall a few fundamental results there reported.

Consider the digraph associated to the Laplacian matrix $\mathrm{L}+\mathrm{L}^u$ and calculate the exclusive parts of the maximal reachable sets of the graph, i.e., $\mathcal{H}_i$ with $i=1,\ldots,\mu$, where $\mu$ is the number of zero eigenvalues of $\mathrm{L}+\mathrm{L}^u$, and the union of the common parts, i.e., $\mathcal{C}$,. Then, let us indicate with $n_i$ the cardinality of $\mathcal{H}_i$, i.e., $n_i=|\mathcal{H}_i|$, and with $v_{1,i}, v_{2,i}, \ldots, v_{n_i,i}$ the nodes belonging to $\mathcal{H}_i$ and adopt a similar notation for $\mathcal{C}$. We define the following permutation matrix

\begin{equation}
\label{eq:Tmatrix}
\mathrm{T}=[e_{v_{1,1}} e_{v_{2,1}} \ldots e_{v_{n_1,1}} ... e_{v_{\mu,1}} \ldots e_{v_{\mu,n_\mu}} e_{v_{C,1}} \ldots e_{v_{C,n_C}}]
\end{equation}

\noindent that, according to \cite{caughman2006kernels}, leads to the following block decomposition for the Laplacian matrix:

\begin{equation}
\label{eq:Ldecomposition}
\begin{array}{l}
\tilde{\mathrm{L}} \triangleq \mathrm{T}^T(\mathrm{L}+\mathrm{L}^u)\mathrm{T}=\\
=\left [
\begin{array}{ccccc}
\mathrm{L}_1 & 0_{n_1\times n_2} & \ldots & 0_{n_1\times n_\mu} & 0_{n_1\times n_C}\\
0_{n_2\times n_1} & \mathrm{L}_2 & \ldots & 0_{n_2\times n_\mu} & 0_{n_2\times n_C}\\
\vdots & & & & \vdots\\
0_{n_\mu\times n_1} & 0_{n_\mu\times n_2} & \ldots & \mathrm{L}_\mu & 0_{n_\mu\times n_C}\\
\mathrm{M}_1 & \mathrm{M}_2 & \ldots & \mathrm{M}_\mu & \mathrm{M}
\end{array}
\right ]
\end{array}
\end{equation}

\noindent where the blocks $\mathrm{L}_i$ with $i=1,\ldots,\mu$ are Laplacian matrices associated with the exclusive parts of the maximal reachable sets of the graph. From the remaining blocks we calculate the vectors $\gamma_i$ with $i=1,\ldots,\mu$ as the solutions of:

\begin{equation}
\label{eq:MiMgammai}
\mathrm{M}_i 1_{n_i}+\mathrm{M}\gamma_i=0
\end{equation}

The partition $\pi^*$ is obtained by considering $\mu$ cells, each associated to one of the exclusive parts of the maximal reachable sets of the graph, $\mathcal{H}_i$, and further cells obtained by grouping together the elements of $\mathcal{C}$ having equal components of the vectors $\gamma_i$, with $i=1,\ldots,\mu$. In this way the set $\mathcal{C}$ is divided into $k$ distinct cells, that is $\mathcal{C} = \bigoplus \limits_{h=1}^{k} \mathcal{C}_h$.

The partition $\pi^*$ has an important property (Corollary 1 of \cite{monaco2019multi}).

\begin{lemma}
\label{lem:coarsest}
If $\mu>1$, then the partition $\pi^*$ is the non-trivial coarsest EEP of the network with Laplacian matrix $\mathrm{L}+\mathrm{L}^u$. If $\mu=1$, then $\pi^*$ coincides with the trivial partition with all nodes grouped in one cell.
\end{lemma}

Clearly, if $\mu=1$, then the multi-consensus reduces to classical consensus, as the digraph is rooted out-branching (see also Remark~\ref{rem:remark33}).

\begin{remark}
If $\mu>1$, then the number of cells in $\pi^*$, indicating the degree of coarseness of the partition and labeled as $n_{\pi^*}$ is such that: $$\mu+1\leq n_{\pi^*} \leq \mu + |\mathcal{C}|=\mu + N-\sum\limits_{i=1}^N{n_i}.$$
\end{remark}

The next Lemma, readapted from Corollary 2 of \cite{monaco2019multi}, formalizes the fact that a multi-agent system reaches the multi-consensus associated with the partition $\pi^*$.

\begin{lemma}
\label{lem:lemmaconvergenzapistar}
A multi-agent system of the form
\begin{equation}
\label{eq:MAScontrolled}
\dot{x}_i(t)=-\sum\limits_{j=1}^N(\mathrm{L}_{ij}+\mathrm{L}_{ij}^u)x_j(t)
\end{equation}

\noindent achieves multi-consensus with respect to groups of nodes that coincide with the cells of the EEP $\pi^*$.
\end{lemma}

We are now ready to state our main result on the design of controllers for multi-consensus of single integrators.

\begin{theorem}
\label{th:th1}
Given the multi-agent system

\begin{equation}
\label{eq:MASnocontrol}
\dot{x}_i(t)=-\sum\limits_{j=1}^N\mathrm{L}_{ij}x_j(t)+u_i(t)
\end{equation}

\noindent and a desired multi-consensus associated with the partition $\pi^Q$ such that each cell contains at most a single rooted node, then it is possible to find controllers (\ref{eq:controllers}) such that the controlled multi-agent system

\begin{equation}
\label{eq:MAScontrolled}
\dot{x}_i(t)=-\sum\limits_{j=1}^N(\mathrm{L}_{ij}+\mathrm{L}_{ij}^u)x_j(t)
\end{equation}

\noindent reaches the desired multi-consensus.
\end{theorem}

\emph{Proof:} Lemma~\ref{lem:lemma1} guarantees that it is possible to find $\mathrm{L}^u$ such that $\pi^Q$ is an EEP for the network with Laplacian $\mathrm{L}+\mathrm{L}^u$. Let us then consider $\pi^*$. By Lemma~\ref{lem:lemmaconvergenzapistar}, we have that the multi-agent system (\ref{eq:MAScontrolled}) achieves a multi-consensus with groups of nodes coinciding with the cells of $\pi^*$. However, as in each cell of $\pi^Q$ there is at most a single rooted node and $\pi^*$, by Lemma~\ref{lem:coarsest}, is the coarsest EEP of the underlying network of the controlled multi-agent system, then each cell of $\pi^Q$ coincides with a cell of $\pi^Q$ or is entirely contained in a cell of $\pi^Q$; hence, either $\pi^Q$ coincides with $\pi^*$ or $\pi^Q$ is finer than $\pi^*$. In both cases, the desired multi-consensus (\ref{eq:multiconsensus}) is reached. Eventually, if $\pi^Q$ is finer than $\pi^*$, then the multi-consensus reached by the controlled multi-agent system will be characterized by two or more cells of $\pi^Q$ which are merged together in a cell of $\pi^*$. $\square$

\begin{remark}
To solve the multi-consensus problem as in Theorem~\ref{th:th1}, the controllers in Eq.~(\ref{eq:controllers}) are designed such that the Laplacian matrix $\mathrm{L}+\mathrm{L}^u$ satisfies Eq.~(\ref{eq:permute}), with the matrix $\mathrm{L}^u$ obtained by solving the integer linear programming problem defined in Lemma~\ref{lem:lemma1} or using the algorithm of Remark~\ref{rem:constructivealgo}. Following this procedure, the minimum number of controllers is used to solve the task.
\end{remark}

%
%
%

\subsection{Numerical examples}
\label{sec:esempiIorder}

\begin{figure}
\begin{center}
\subfigure[]{\includegraphics[width=0.24\textwidth]{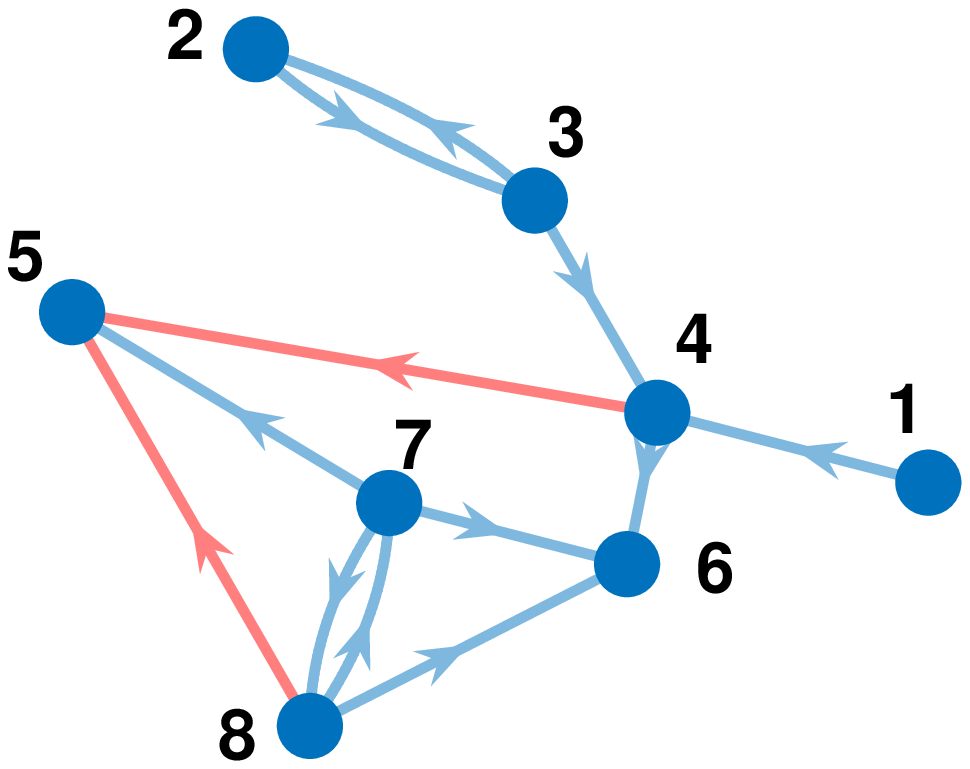}}
\subfigure[]{\includegraphics[width=0.24\textwidth]{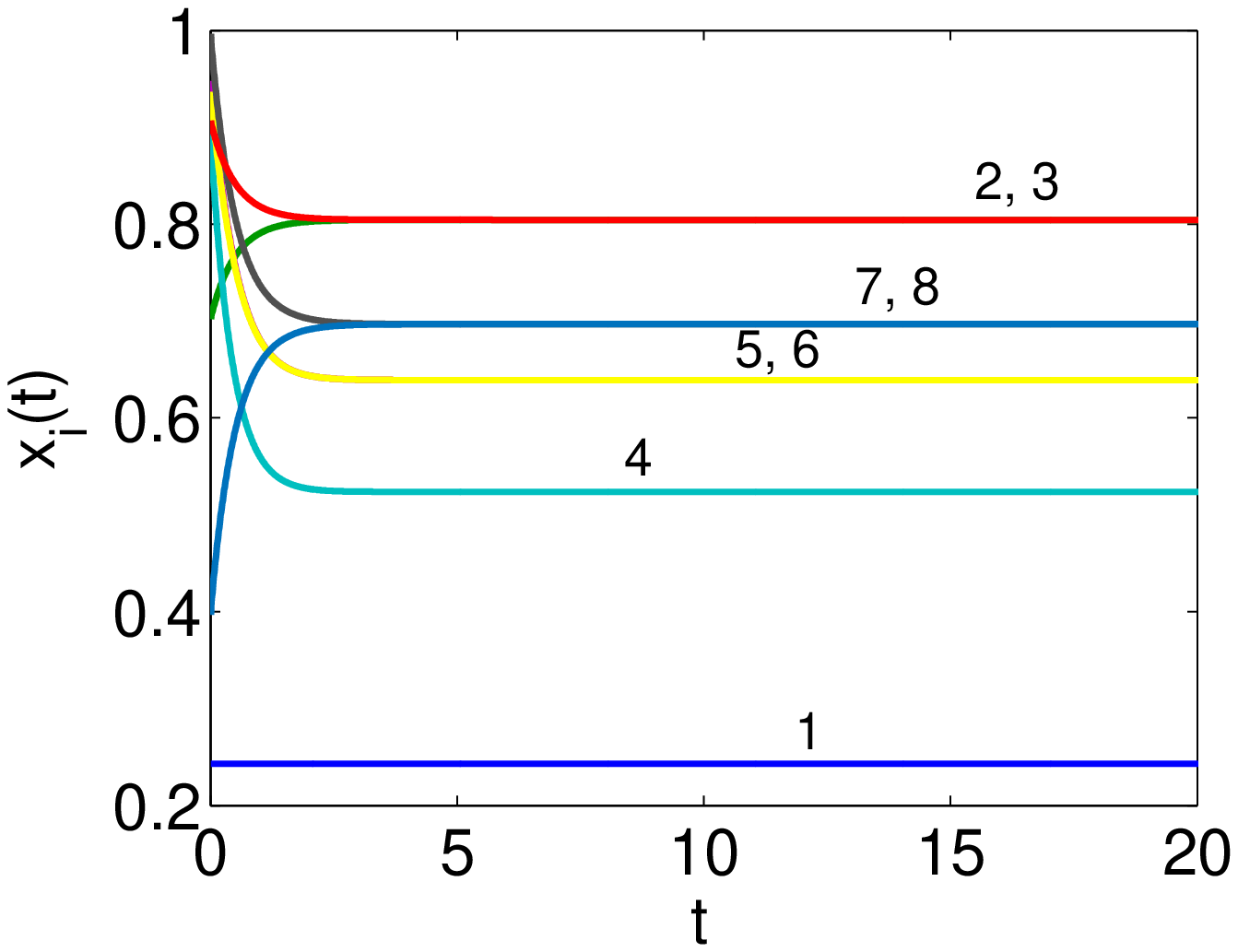}}
\caption{\label{fig:graph1} Multi-consensus of single integrators. (a) Digraph with $N=8$ modeling the interactions among the agents. In blue the links of the original graph are shown, in red those added to reach the desired multi-consensus discussed in Example~1. (b) Time evolution of variables $\mathbf{x}_i(t)$.}
\end{center}
\end{figure}

\emph{Example 1}. As a first example of multi-consensus for single integrators, we consider a multi-agent system~(\ref{eq:MAS}) with interaction topology given by the digraph reported in Fig.~\ref{fig:graph1}(a) (blue lines) and suppose that in the target multi-consensus the agents are grouped in the following clusters $C_1=\{2,3\}$, $C_2=\{5,6\}$, $C_3=\{7,8\}$, $C_4=\{1\}$, $C_5=\{4\}$, i.e., $\pi^Q=\{C_1,C_2,C_3,C_4,C_5\}$. Solving the integer linear programming problem of Lemma~\ref{lem:lemma1}, we obtain that two links have to be added to the original structure, i.e., link $(4,5)$ and $(8,5)$. In Fig.~\ref{fig:graph1}(a) these links have been superimposed in red to the original structure of the network.

The matrix $\mathrm{P}_H$ associated to the partition $\pi^Q=\{C_1,C_2,C_3,C_4,C_5\}$ is given by:

\begin{equation}
\label{PhExample1}
\begin{array}{l}
\mathrm{P}_H=
\left (
\begin{array}{cccccccc}
1 & 0 & 0 & 0 & 0 & 0 & 0 & 0 \\
0 & 0.5 & 0.5 & 0 & 0 & 0 & 0 & 0 \\
0 & 0.5 & 0.5 & 0 & 0 & 0 & 0 & 0 \\
0 & 0 & 0 & 1 & 0 & 0 & 0 & 0 \\
0 & 0 & 0 & 0 & 0.5 & 0.5 & 0 & 0 \\
0 & 0 & 0 & 0 & 0.5 & 0.5 & 0 & 0 \\
0 & 0 & 0 & 0 & 0 & 0 & 0.5 & 0.5 \\
0 & 0 & 0 & 0 & 0 & 0 & 0.5 & 0.5 \\
\end{array}
\right )
\end{array}
\end{equation}

\noindent while the Laplacian matrix associated to the resulting digraph in Fig.~\ref{fig:graph1}(a) is given by:

\begin{equation}
\label{LLuExample1}
\begin{array}{l}
\mathrm{L}+\mathrm{L}^u=
\left (
\begin{array}{cccccccc}
0 & 0 & 0 & 0 & 0 & 0 & 0 & 0 \\
0 & 1 & -1 & 0 & 0 & 0 & 0 & 0 \\
0 & -1 & 1 & 0 & 0 & 0 & 0 & 0 \\
-1 & 0 & -1 & 2 & 0 & 0 & 0 & 0 \\
0 & 0 & 0 & -1 & 3 & 0 & -1 & -1 \\
0 & 0 & 0 & -1 & 0 & 3 & -1 & -1 \\
0 & 0 & 0 & 0 & 0 & 0 & 1 & -1 \\
0 & 0 & 0 & 0 & 0 & 0 & -1 & 1 \\
\end{array}
\right )
\end{array}
\end{equation}

Direct calculation shows that the matrices $\mathrm{P}_H$ (\ref{PhExample1}) and  $\mathrm{L}+\mathrm{L}^u$  (\ref{LLuExample1}) satisfy eq. (\ref{eq:permute})
such that $\pi^Q=\{C_1,C_2,C_3,C_4,C_5\}$ is an EEP for the digraph in Fig.~\ref{fig:graph1}(a).

For the resulting digraph we now calculate the partition $\pi^*= \{\mathcal{H}_i, \mathcal{C}_h\}$. The reachable sets are $\mathcal{H}_i= \{\mathcal{H}_1, \mathcal{H}_2, \mathcal{H}_3\}=\{\{1\},\{2,5\},\{7,8\}\}$, while the union of the common parts is $\mathcal{C}=\{4,5,6\}$.
The permutation matrix $\mathrm{T}$~(\ref{eq:Tmatrix}) is thus given by $\mathrm{T}=[e_{1}~e_{2}~ e_{3}~e_{7}~e_{8}~e_{4}~e_{5}~e_{6}]$, leading to the following block-decomposed Laplacian:

\begin{equation}
\label{eq:laplacianEx1}
\begin{array}{l}
\tilde{\mathrm{L}}=
\left (
\begin{array}{c|cc|cc|ccc}
0 & 0 & 0 & 0 & 0 & 0 & 0 & 0 \\\hline
0 & 1 & -1 & 0 & 0 & 0 & 0 & 0 \\
0 & -1 & 1 & 0 & 0 & 0 & 0 & 0 \\\hline
0 & 0 & 0 & 1 & -1 & 0 & 0 & 0 \\
0 & 0 & 0 & -1 & 1 & 0 & 0 & 0 \\\hline
-1 & 0 & -1 & 0 & 0 & 2 & 0 & 0 \\
0 & 0 & 0 & -1 & -1 & -1 & 3 & 0 \\
0 & 0 & 0 & -1 & -1 & -1 & 0 & 3 \\
\end{array}
\right )
\end{array}
\end{equation}

\noindent where the lines suggest the division into blocks corresponding to the reachable sets $\mathcal{H}_i$. The last block is the one related to the common part $\mathcal{C}$. Solving Eqs.~(\ref{eq:MiMgammai}), we obtain that $\mathcal{C}=\{\mathcal{C}_1, \mathcal{C}_2\}=\{\{4\},\{5,6\}\}$, such that $\pi^*=\{H_1,H_2,H_3,C_1,C_2\}$, from which we derive that, in this case, the partition $\pi^*$ and $\pi^Q$ coincide. Numerical simulations of the controlled multi-agent system~(\ref{eq:MAScontrolled}) from random initial conditions confirm that the system achieves the desired multi-consensus. An illustrative trajectory is shown in Fig.~\ref{fig:graph1}(b), with the different clusters reaching different values of the consensus.

\begin{figure}
\begin{center}
\subfigure[]{\includegraphics[width=0.24\textwidth]{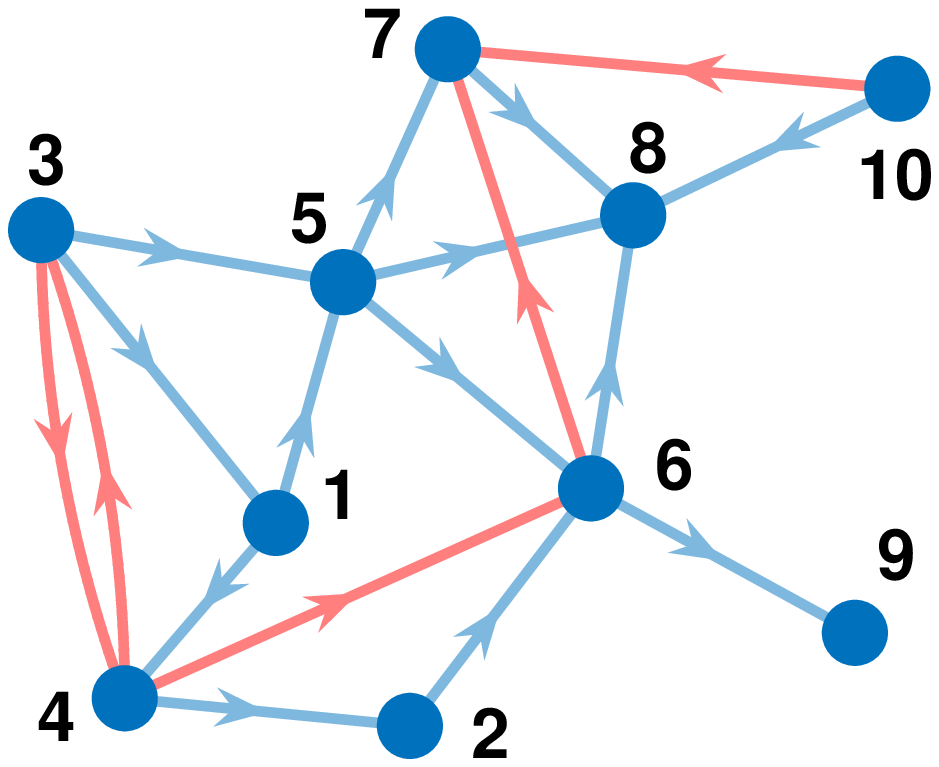}}
\subfigure[]{\includegraphics[width=0.24\textwidth]{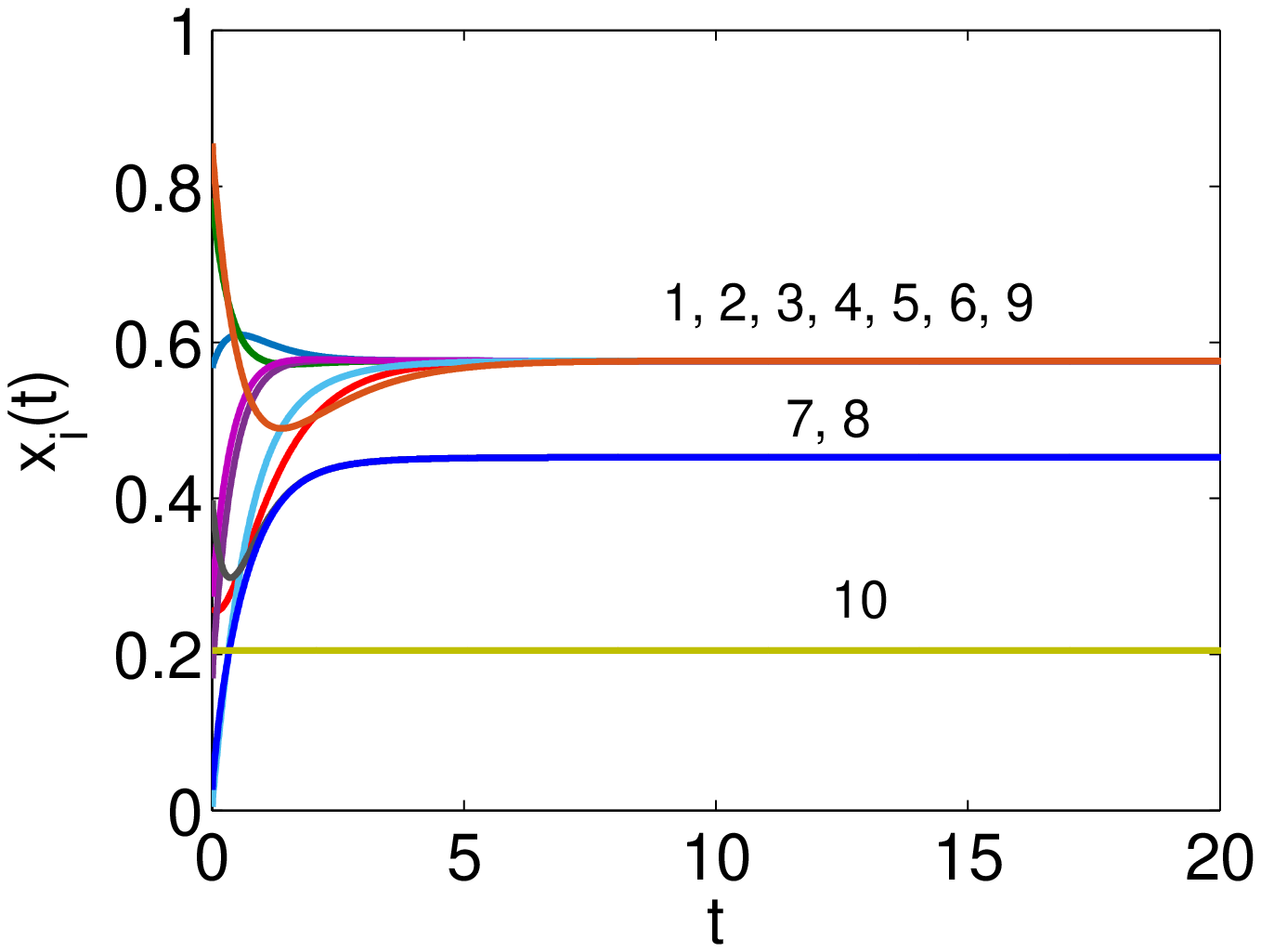}}
\caption{\label{fig:graph2} Multi-consensus of single integrators. (a) Digraph with $N=10$ modeling the interaction among agents. In blue the links of the original graph are shown, in red those added to reach the desired multi-consensus discussed in Example~2. (b) Time evolution of variables $\mathbf{x}_i(t)$.}
\end{center}
\end{figure}

\emph{Example 2}. As second example, we consider the multi-agent system with the digraph reported in Fig.~\ref{fig:graph2}(a) and the target multi-consensus defined by the following partition $\pi^Q= \{\{1,4\},\{2,3\},\{5,6\},\{7,8\},\{9\},\{10\}\}$. The solution of the optimization problem in Lemma~\ref{lem:lemma1} yields the addition of 5 links: $(3,4)$, $(4,3)$, $(4,6)$, $(6,7)$, and $(10,7)$ superimposed in Fig.~\ref{fig:graph2} to the original structure (the links representing the controllers are shown in red, while those of the original structure are in blue). In this example, the matrices $\mathrm{P}_H$ and  $\mathrm{L}+\mathrm{L}^u$ are given by:

\begin{equation}
\begin{array}{l}
\mathrm{P}_H=
\left (
\begin{array}{cccccccccc}
 0.5 & 0 & 0 & 0.5 & 0 & 0 & 0 & 0 & 0 & 0 \\
 0 & 0.5 & 0.5 & 0 & 0 & 0 & 0 & 0 & 0 & 0 \\
 0 & 0.5 & 0.5 & 0 & 0 & 0 & 0 & 0 & 0 & 0 \\
 0.5 & 0 & 0 & 0.5 & 0 & 0 & 0 & 0 & 0 & 0 \\
 0 & 0 & 0 & 0 & 0.5 & 0.5 & 0 & 0 & 0 & 0 \\
 0 & 0 & 0 & 0 & 0.5 & 0.5 & 0 & 0 & 0 & 0 \\
 0 & 0 & 0 & 0 & 0 & 0 & 0.5 & 0.5 & 0 & 0 \\
 0 & 0 & 0 & 0 & 0 & 0 & 0.5 & 0.5 & 0 & 0 \\
 0 & 0 & 0 & 0 & 0 & 0 & 0 & 0 & 1 & 0 \\
 0 & 0 & 0 & 0 & 0 & 0 & 0 & 0 & 0 & 1 \\
\end{array}
\right )
\end{array}
\end{equation}

\noindent and

\begin{equation}
\begin{array}{l}
\mathrm{L}+\mathrm{L}^u=
\left (
\begin{array}{cccccccccc}
 1    &     0 &  -1  &       0  &       0    &     0    &     0    &     0   &      0    &     0 \\
 0   & 1    &     0  & -1    &     0    &     0   &      0    &     0      &   0      &   0 \\
 0   &      0  &  1 &  -1   &      0  & 0    &     0 &  0    &     0 &  0 \\
 -1   &      0 &  -1  &  2  &  0     &    0  &  0     &    0     &    0      &   0 \\
 -1    &     0  & -1   &      0  &  2   &      0   &      0    &     0   &      0   &      0 \\
 0  & -1  & 0  & -1  & -1 &   3    &   0 & 0    &     0    &     0 \\
 0  &  0   &      0    &     0  & -1  & -1  &  3   &      0   &      0  & -1 \\
 0  &  0    &    0      &   0  & -1  & -1 &  -1 &  4   &   0 &  -1 \\
 0  &       0   &      0    &     0    &    0 &  -1   &      0   &      0  &  1  &       0 \\
 0   &      0     &    0   &      0     &    0    &     0   &      0   &      0   &      0   &      0 \\
\end{array}
\right )
\end{array}
\end{equation}

Direct calculation shows that they satisfy eq. (\ref{eq:permute})
such that $\pi^Q= \{\{1,4\},\{2,3\},\{5,6\},\{7,8\},\{9\},\{10\}\}$ is an EEP for the digraph in Fig.~\ref{fig:graph2}(a).

By applying the permutation matrix $\mathrm{T}=[e_{1}~e_{2}~ e_{3}~e_{4}~e_{5}~e_{6}~e_{9}~e_{10}~e_{7}~e_{8}]$, we obtain the block-decomposed Laplacian:

\begin{equation}
\label{eq:laplacianEx2}
\begin{array}{l}
\tilde{\mathrm{L}}=\\
=\left (
\begin{array}{ccccccc|c|cc}
1 & 0 & -1 & 0 & 0 & 0 & 0 & 0 & 0 & 0\\
0 & 1 & 0 & -1 & 0 & 0 & 0 & 0 & 0 & 0\\
0 & 0 & 1 & -1 & 0 & 0 & 0 & 0 & 0 & 0\\
-1 & 0 & -1 & 2 & 0 & 0 & 0 & 0 & 0 & 0\\
-1 & 0 & -1 & 0 & 2 & 0 & 0 & 0 & 0 & 0\\
0 & -1 & 0 & -1 & -1 & 3 & 0 & 0 & 0 & 0\\
0 & 0 & 0 & 0 & 0 & -1 & 1 & 0 & 0 & 0\\\hline
0 & 0 & 0 & 0 & 0 & 0 & 0 & 0 & 0 & 0\\\hline
0 & 0 & 0 & 0 & -1 & -1 & 0 & -1 & 3 & 0\\
0 & 0 & 0 & 0 & -1 & -1 & 0 & -1 & -1 & 4\\
\end{array}
\right )
\end{array}
\end{equation}

The matrix is partitioned in three main blocks, corresponding to the reachable sets and to the common part of the graph: $\pi^*=\{\mathcal{H}_1, \mathcal{H}_2, \mathcal{C}\}= \{\{1,2,3,4,5,6,9\}, \{10\}, \{7,8\} \}$. In this case the partition $\pi^Q$ is contained in $\pi^*$, as four of the original clusters merged together in the cell $\{\mathcal{H}_1\}$. As a result, the units of the multi-agent system ~(\ref{eq:MAScontrolled}) converge to three different values, as shown in Fig.~\ref{fig:graph2}(b).


\section{Control of multi-consensus of second-order dynamics}
\label{sec:secondOrder}

\subsection{Controller design}

Before discussing the design of the controllers for the case of second-order dynamics, we define a matrix $\mathrm{R} \in \mathbb{R}^{N\times N}$ associated to a generic partition $\pi$ as a block-diagonal matrix where each block is $\mathrm{R}_i=\mathrm{I}_{n_i}-\frac{1}{n_i}\mathbf{1}_{n_i}\mathbf{1}_{n_i}^T$, with $n_i=|C_i|$, where $C_i$ with $i=1, \ldots, M$ are the cells of the partition $\pi$. Equivalently, $\mathrm{R}$ can be defined as $\mathrm{R}=\mathrm{I}_N-\mathrm{P}_H$, where $\mathrm{P}_H$ is the projection operator associated to the partition $\pi$.

In the following lemma we demonstrate a property of the eigenvalues of the product of this matrix and the Laplacian $\mathrm{L}$.


\begin{lemma}
\label{lem:eigproduct}
Consider a graph and an EEP $\pi:\{C_1,C_2,\ldots,C_M\}$. Also consider the matrix $\mathrm{R}=(\mathrm{I}_N-\mathrm{P}_H)$ associated to $\pi$ and the Laplacian matrix $\mathrm{L}$ of the graph. Then, the eigenvalues of $\mathrm{R}\mathrm{L}$ are given by:
\begin{equation}
\{\lambda_i(\mathrm{R}\mathrm{L})\}=\{\lambda_i(\mathrm{L})\}
\setminus \{\lambda_i(\mathrm{L}_{\pi})\}\cup \mathcal{O}
\end{equation}
\noindent where $\mathrm{L}_{\pi}$ is the Laplacian of the quotient graph and $\mathcal{O}$ a set of $M$ zeros.
\end{lemma}

\emph{Proof:} Consider the matrix $\mathrm{P}_H$ associated with the partition $\pi$. This matrix has eigenvalues 1 with multiplicity $M$ and 0 with multiplicity $N-M$. Let us order them as follows $\lambda_1=\ldots=\lambda_M=1$ and $\lambda_{M+1}=\ldots=\lambda_N=0$. The matrix is symmetric, and thus there exists an orthogonal matrix $\bar{\mathrm{T}}$ such that $\bar{\mathrm{T}}^T\mathrm{P}_H\bar{\mathrm{T}}=\mathrm{diag}\{\lambda_1,\lambda_2,\ldots,\lambda_N\}$.
Following \cite{gambuzza2019criterion}, the matrix $\bar{\mathrm{T}}$ is rewritten as:

\begin{equation}
\label{eq:matrixT}
\bar{\mathrm{T}}=\left [ \begin{array}{ccccc} \bar{\mathrm{T}}^{(1)}_{1} & \bar{\mathrm{T}}^{(0)}_{1} & 0 & \ldots & 0 \\
\bar{\mathrm{T}}^{(1)}_{2} & 0 & \bar{\mathrm{T}}^{(0)}_{2} & \ldots & 0 \\
\vdots & & & \ddots & \vdots \\
\bar{\mathrm{T}}^{(1)}_{M} & 0 & 0 & \ldots & \bar{\mathrm{T}}^{(0)}_{M}
\end{array}\right] \end{equation}

\noindent where each block $\bar{\mathrm{T}}^{(0)}_{l} \in \mathbb{R}^{n_l \times (n_l-1)}$ with $l=1,\ldots,M$ is such that $\bar{\mathrm{T}}^{(0),T}_{l} (\frac{1}{n_l}\mathrm{1}_{n_l}\mathrm{1}_{n_l}^T)\bar{\mathrm{T}}^{(0)}_{l}=0\cdot\mathrm{I}_{n_l-1}$ and $\bar{\mathrm{T}}^{(0),T}_{l}\bar{\mathrm{T}}^{(0)}_{l}=\mathrm{I}_{n_l-1}$, i.e., it contains the $n_l$ orthogonal eigenvectors associated to the eigenvalue $\lambda=0$ of the $l$-th block appearing in $\mathrm{P}_H$. In the block $\bar{\mathrm{T}}^{(1)}_{l} \in \mathbb{R}^{n_l\times M}$ the $l$-th column is the eigenvector associated to the eigenvalue $\lambda=1$ of the $l$-th block appearing in $\mathrm{P}_H$, while all the other columns are zeros. By direct calculation, we obtain that:

\begin{equation}
\bar{\mathrm{T}}^T\mathrm{P}_H\bar{\mathrm{T}}=\left [ \begin{array}{cc} 1\cdot \mathrm{I}_{M} & \\ & 0\cdot \mathrm{I}_{N-M}
\end{array}
\right]
\end{equation}

Similarly, one derives that:

\begin{equation}
\label{eq:formaR}
\bar{\mathrm{T}}^T\mathrm{R}\bar{\mathrm{T}}=\left [ \begin{array}{cc} 0\cdot \mathrm{I}_{M} & \\ & 1\cdot \mathrm{I}_{N-M}
\end{array}
\right]
\end{equation}

Consider now $\bar{\mathrm{T}}^T\mathrm{L}\bar{\mathrm{T}}$ and partition it conformly to (\ref{eq:formaR}) as $\bar{\mathrm{T}}^T\mathrm{L}\bar{\mathrm{T}}=\left [ \begin{array}{cc} \bar{\mathrm{L}}_{11} & \bar{\mathrm{L}}_{11} \\ \bar{\mathrm{L}}_{21} & \bar{\mathrm{L}}_{22}\end{array}
\right]$. Consequently, we have:

\begin{equation}
\begin{array}{l}
\bar{\mathrm{T}}^T\mathrm{R}\mathrm{L}\bar{\mathrm{T}}=\bar{\mathrm{T}}^T\mathrm{R}\bar{\mathrm{T}}\bar{\mathrm{T}}^T\mathrm{L}\bar{\mathrm{T}}=\\
=\left [ \begin{array}{cc} 0\cdot \mathrm{I}_{M} & \\ & 1\cdot \mathrm{I}_{N-M}
\end{array}
\right]\left [ \begin{array}{cc} \bar{\mathrm{L}}_{11} & \bar{\mathrm{L}}_{11} \\ \bar{\mathrm{L}}_{21} & \bar{\mathrm{L}}_{22}
\end{array}
\right]=\\
=\left [ \begin{array}{cc} 0\cdot \mathrm{I}_{M} & \mathrm{O}_{M,N-M} \\ \bar{\mathrm{L}}_{21} & \bar{\mathrm{L}}_{22}
\end{array}
\right]
\end{array}
\end{equation}

Since $\bar{\mathrm{T}}^T\mathrm{R}\mathrm{L}\bar{\mathrm{T}}$ and $\mathrm{R}\mathrm{L}$ have the same eigenvalues, we conclude that the eigenvalues of $\mathrm{R}\mathrm{L}$ are those of $\bar{\mathrm{L}}_{22}$ along with $M$ zero eigenvalues.

We now study the eigenvalues of $\bar{\mathrm{L}}_{22}$. Let us rewrite the matrix $\bar{\mathrm{T}}$ as
$\bar{\mathrm{T}}=[\bar{\mathrm{T}}^{(1)} | \bar{\mathrm{T}}^{(0)}]$ where $\bar{\mathrm{T}}^{(1)}=[\bar{\mathrm{T}}_1^{(1),T} \bar{\mathrm{T}}_2^{(1),T} \ldots \bar{\mathrm{T}}_M^{(1),T}]^T$ and $\bar{\mathrm{T}}^{(0)}=\mathrm{diag}\{\bar{\mathrm{T}}_1^{(0)} \ldots \bar{\mathrm{T}}_M^{(0)} \}$. Following \cite{ji2007graph}, we can select $\bar{\mathrm{T}}^{(1)}$ as $\bar{\mathrm{T}}^{(1)}=\mathrm{P}(\mathrm{P}^T\mathrm{P})^{-\frac{1}{2}}$. Then, we have that

\begin{equation}
\begin{array}{l}
\bar{\mathrm{T}}^T\mathrm{L}\bar{\mathrm{T}}=\left [ \begin{array}{cc}   \bar{\mathrm{T}}^{(1),T}\mathrm{L}\bar{\mathrm{T}}^{(1)} &  \bar{\mathrm{T}}^{(1),T}\mathrm{L}\bar{\mathrm{T}}^{(0)} \\
\bar{\mathrm{T}}^{(0),T}\mathrm{L}\bar{\mathrm{T}}^{(1)} & \bar{\mathrm{T}}^{(0),T}\mathrm{L}\bar{\mathrm{T}}^{(0)}
\end{array}
\right]
\end{array}
\end{equation}

\noindent and so $\bar{\mathrm{L}}_{11}=\bar{\mathrm{T}}^{(1),T}\mathrm{L}\bar{\mathrm{T}}^{(1)}$. This term can be further manipulated by substituting $\bar{\mathrm{T}}^{(1)}=\mathrm{P}(\mathrm{P}^T\mathrm{P})^{-\frac{1}{2}}$:

\begin{equation}
\begin{array}{lll}
\bar{\mathrm{L}}_{11}&=&(\mathrm{P}^T\mathrm{P})^{-\frac{1}{2}}\mathrm{P}^T\mathrm{L}\mathrm{P}(\mathrm{P}^T\mathrm{P})^{-\frac{1}{2}}=\\
& = & (\mathrm{P}^T\mathrm{P})^{\frac{1}{2}}(\mathrm{P}^T\mathrm{P})^{-1}\mathrm{P}^T\mathrm{L}\mathrm{P}(\mathrm{P}^T\mathrm{P})^{-\frac{1}{2}}=\\
& = & (\mathrm{P}^T\mathrm{P})^{\frac{1}{2}}\mathrm{L}_\pi(\mathrm{P}^T\mathrm{P})^{-\frac{1}{2}}
\end{array}
\end{equation}

Hence, $\bar{\mathrm{L}}_{11}$ is similar to $\mathrm{L}_\pi$ and, so, has the same eigenvalues. It follows that the eigenvalues of $\bar{\mathrm{L}}_{22}$ are those eigenvalues of $\mathrm{L}$ which are not of $\mathrm{L}_\pi$. From this, the thesis immediately follows. $\square$

For the multi-agent system with second-order dynamics (\ref{eq:IIorderMatrixForm}), the design of the controllers consists of two steps. The first step is analogous to the case of single integrator dynamics: Lemma~\ref{lem:lemma1} is applied to find $\mathrm{L}^u$ such that $\pi^Q$ is an EEP for the network underlying the multi-agent system. In the second step, the gains $k_1$ and $k_2$ are selected. These two steps are summarized in the following theorem.

\begin{theorem}
\label{th:mainforthesecondorder}
Given the multi-agent system

\begin{equation}
\label{eq:IIorderMatrixForm}
\dot{\mathbf{x}}_i(t)= \mathrm{A} \mathbf{x}_i(t) - \sum\limits_{j=1}^N\mathrm{L}_{ij} \mathrm{B}\mathrm{K}\mathbf{x}_j(t) + \mathrm{B} \mathbf{u}_i(t)
\end{equation}

\noindent and a desired multi-consensus associated with the partition $\pi^Q$ such that each cell contains at most a single rooted node, then it is possible to find controllers (\ref{eq:controllersIIorder}) such that for the controlled multi-agent system

\begin{equation}
\label{eq:IIorderMatrixFormControlled}
\dot{\mathbf{x}}_i(t)= \mathrm{A} \mathbf{x}_i(t) - \sum\limits_{j=1}^N(\mathrm{L}_{ij} +\mathrm{L}^u_{ij} )\mathrm{B}\mathrm{K}\mathbf{x}_j(t)
\end{equation}

\noindent the desired multi-consensus manifold exists. In addition,
let $\gamma$ be the smallest non-zero element of the set $\{\lambda_i(\mathrm{L}+\mathrm{L}^u)\}
\setminus \{\lambda_i((\mathrm{L}+\mathrm{L}^u)_{\pi^*})\}\cup \mathcal{O}$, with $\pi^*:\{\mathcal{H}_1,\mathcal{H}_2,\ldots,\mathcal{H}_\mu,\mathcal{C}_1,\ldots,\mathcal{C}_h\}$ ($\mathcal{H}_i$ with $i=1,\ldots,\mu$ are the exclusive parts of the reachable sets of the digraph obtained considering the original and the control layer and $\mathcal{C}_i$ with $i=1,\ldots,h$ the cells in which the common part $\mathcal{C}$ is splitted) and where $\mathcal{O}$ is a set of $\mu+h$ zeros, then if $k_1>\frac{a}{\gamma}$ and $k_2>\frac{b}{\gamma}$, the desired multi-consensus is stable.
\end{theorem}

\emph{Proof:} We first apply Lemma~\ref{lem:lemma1} to find $\mathrm{L}^u$ such that $\pi^Q$ is an EEP for the network underlying the controlled multi-agent system. Then, the EEP $\pi^*$ is considered. Analogously, to Theorem~\ref{th:th1}, it can be shown that $\pi^Q$ either coincides with $\pi^*$ or is a finer partition than it. This proves that for the controlled multi-agent system the desired multi-consensus state exists. Now, we show that this state is stable.

To this aim, the compact form of the controlled multi-agent system is taken into account:

\begin{equation}
\label{eq:compactForm}
\dot{\mathbf{X}}(t) = (\mathrm{I}_N \otimes \mathrm{A} - (\mathrm{L} + \mathrm{L}^u) \otimes \mathrm{B}\mathrm{K}) \mathbf{X}(t)
\end{equation}

\noindent where $\mathbf{X}=[\mathbf{x}_1^T,\mathbf{x}_2^T,\ldots,\mathbf{x}_N^T]^T$ is the stack vector of the state variables of the agents.  The nodes of the multi-agent system are then permuted such that the Laplacian in the new reference system has the block decomposition (\ref{eq:Ldecomposition}). This is equivalent to consider new state variables $\mathbf{Y}(t)=(\mathrm{T}^T\otimes \mathrm{I}_2)\mathbf{X}(t)$ where $\mathrm{T}$ is the permutation matrix given by~(\ref{eq:Tmatrix}). Then, the system dynamics in the new reference system reads:

\begin{equation}
\label{eq:compactForm2}
\dot{\mathbf{Y}}(t) = (\mathrm{I}_N \otimes \mathrm{A} - \tilde{\mathrm{L}}  \otimes \mathrm{B}\mathrm{K}) \mathbf{Y}(t)
\end{equation}

We now define new variables through the matrix $\mathrm{R}$ as follows:

\begin{equation}
\label{eq:z}
\mathbf{Z}(t)=(\mathrm{R}\otimes \mathrm{I}_2)\mathbf{Y}(t)
\end{equation}

\noindent representing the errors with respect to the mean value of the position and velocity state variables, i.e., $\mathbf{z}_i(t)=\mathbf{y}_i(t)-\bar{\mathbf{y}}_h(t)$ inside each cluster $C_h$, where $\bar{\mathbf{y}}_h(t)=\frac{1}{n_h}\sum \limits_{j \in C_h} \mathbf{y}_j(t)$. Notice, in fact, that the matrix $\mathrm{R}$ represents an orthogonal projection onto the multi-consensus manifold, enabling the study of the stabilization effect of the control law on the distance of the single agent states from the consensus value of each cluster. We have that

\begin{equation}
\label{eq:ErrorEq}
\begin{array}{lll}
\dot{\mathbf{Z}}(t) & = & (\mathrm{R}\otimes \mathrm{I}_2)\dot{\mathbf{Y}}(t) =\\
& = & (\mathrm{R}\otimes \mathrm{I}_2)(\mathrm{I}_N \otimes \mathrm{A} - \tilde{\mathrm{L}} \otimes \mathrm{B}\mathrm{K}) \mathbf{Y} =\\
 & = &(\mathrm{R} \otimes \mathrm{A})\mathbf{Y} - (\mathrm{R} \tilde{\mathrm{L}} \otimes \mathrm{B}\mathrm{K}) \mathbf{Y} \\
\end{array}
\end{equation}

From Eq.~(\ref{eq:z}) and the definition of $\mathrm{R}= \mathrm{I}_N-\mathrm{P}_H$, we get $\mathbf{Z}(t)=\mathbf{Y}(t) -(\mathrm{P}_H \otimes \mathrm{I}_2)\mathbf{Y}(t)$, thus $\mathbf{Y}(t)= \mathbf{Z}(t)+ (\mathrm{P}_H \otimes \mathrm{I}_2)\mathbf{Y}(t)$. Substituting this expression in Eq.~(\ref{eq:ErrorEq}) we obtain

\begin{equation}
\label{eq:zetaCompleta}
\begin{array}{lll}
\dot{\mathbf{Z}}(t) &=& (\mathrm{R} \otimes \mathrm{A})\mathbf{Z}(t) +(\mathrm{R} \otimes \mathrm{A})(\mathrm{P}_H \otimes \mathrm{I}_2)\mathbf{Y}(t) \\ & & -(\mathrm{R} \tilde{\mathrm{L}} \otimes \mathrm{B}\mathrm{K}) \mathbf{Z}(t)  \\ & &-(\mathrm{R} \tilde{\mathrm{L}} \otimes \mathrm{B}\mathrm{K})(\mathrm{P}_H \otimes \mathrm{I}_2)\mathbf{Y}(t)
\end{array}
\end{equation}

The second and the last term of the right-hand side of this equation are zero. In fact, for the second term we have that $(\mathrm{R} \otimes \mathrm{A})(\mathrm{P}_H \otimes \mathrm{I}_2)= (\mathrm{R} \mathrm{P}_H) \otimes \mathrm{A}$, but $(\mathrm{R} \mathrm{P}_H)= (\mathrm{I}_N-\mathrm{P}_H)\mathrm{P}_H = \mathrm{P}_H - \mathrm{P}^2_H=0$ since $\mathrm{P}^2_H=\mathrm{P}_H$. For the last term, we have that $(\mathrm{R} \tilde{\mathrm{L}} \otimes \mathrm{B}\mathrm{K})(\mathrm{P}_H \otimes \mathrm{I}_2)= (\mathrm{R} \tilde{\mathrm{L}} \mathrm{P}_H) \otimes (\mathrm{B}\mathrm{K})$, but $(\mathrm{R} \tilde{\mathrm{L}} \mathrm{P}_H) = \tilde{\mathrm{L}}\mathrm{P}_H - \mathrm{P}_H\tilde{\mathrm{L}}\mathrm{P_H}=0$, because of~(\ref{eq:permute}).

Summing up, Eq.~(\ref{eq:zetaCompleta}) becomes

\begin{equation}
\label{eq:zetaCompletaDef}
\begin{array}{lll}
\dot{\mathbf{Z}}(t) &=& (\mathrm{R} \otimes \mathrm{A})\mathbf{Z}(t) - (\mathrm{R} \tilde{\mathrm{L}} \otimes \mathrm{B}\mathrm{K}) \mathbf{Z}(t)
\end{array}
\end{equation}

Considering the block structure of $\tilde{\mathrm{L}}$ as in (\ref{eq:Ldecomposition}) and taking into account that $\mathrm{R}$ can be written as $\mathrm{R}=\mathrm{diag} \{ \mathrm{R}_1,\ldots, \mathrm{R}_{\mu+1} \}$ with $\mathrm{R}_i=\mathrm{I}_{n_i}-\frac{1}{n_i}\mathbf{1}_{n_i}\mathbf{1}_{n_i}^T$ for $i=1,\ldots,\mu$ and $\mathrm{R}_{\mu+1}=\mathrm{diag}\{ \mathrm{I}_{\delta_i}-\frac{1}{\delta_i}\mathbf{1}_{\delta_i}\mathbf{1}_{\delta_i}^T \}$ with $\delta_i=|\mathcal{C}_i|$ and $\delta=|\mathcal{C}|=\sum \delta_i$, then the product matrix $\mathrm{R}\tilde{\mathrm{L}}$ has the following block decomposition, conform to that of $\tilde{\mathrm{L}}$:

\begin{equation}
\label{eq:RLdecomposition}
\begin{array}{l}
\mathrm{R}\tilde{\mathrm{L}}=\\
=\left [
\begin{array}{ccccc}
\mathrm{R}_1\mathrm{L}_1 & 0_{n_1\times n_2} & \ldots & 0_{n_1\times n_\mu} & 0_{n_1\times n_C}\\
0_{n_2\times n_1} & \mathrm{R}_2\mathrm{L}_2 & \ldots & 0_{n_2\times n_\mu} & 0_{n_2\times n_C}\\
\vdots & & & & \vdots\\
0_{n_\mu\times n_1} & 0_{n_\mu\times n_2} & \ldots & \mathrm{R}_\mu \mathrm{L}_\mu & 0_{n_\mu\times n_C}\\
\mathrm{R}_{\mu+1} \mathrm{M}_1 & \mathrm{R}_{\mu+1}\mathrm{M}_2 & \ldots & \mathrm{R}_{\mu+1}\mathrm{M}_\mu & \mathrm{R}_{\mu+1}\mathrm{M}
\end{array}
\right ]
\end{array}
\end{equation}

This yields that Eq.~(\ref{eq:zetaCompletaDef}) can be explicitly rewritten as

\begin{equation}
\label{eq:decoupledEqs}
\dot{\mathbf{Z}}_i(t)= (\mathrm{R}_{n_i} \otimes \mathrm{A}) \mathbf{Z}_i(t) - (\mathrm{R}_i \tilde{\mathrm{L}}_i \otimes \mathrm{B}\mathrm{K}) \mathbf{Z}_i(t) \\
\end{equation}

\noindent with $i=1,\ldots, \mu$ and

\begin{equation}
\label{eq:decoupledEqsRM}
\begin{array}{lll}
\dot{\mathbf{Z}}_{\mu+1}(t) & = & [(\mathrm{R}_{\delta} \otimes \mathrm{A}) - (\mathrm{R}_{\delta} \mathrm{M}_{\delta}\otimes\mathrm{B}\mathrm{K})]\mathbf{Z}_{\mu+1}(t) \\
&& - \sum \limits_ {h=1}^{\mu}( \mathrm{R}_h \mathrm{M}_h \otimes \mathrm{B}\mathrm{K}) \mathbf{Z}_{h}(t)
\end{array}
\end{equation}

\noindent where $\mathbf{Z}_i(t)$ with $i=1,\ldots, \mu$ groups the state variables of the units belonging to the cell $i$, i.e., $\mathbf{Z}_i(t)=[\mathbf{z}_1(t)^T, \mathbf{z}_2(t)^T, \ldots,\mathbf{z}_{n_i}(t)^T]$, and $\mathbf{Z}_{\mu+1}(t)$ those of the common part, i.e., $\mathbf{Z}_{\mu+1}(t)=[\mathbf{z}_1(t)^T, \mathbf{z}_2(t)^T, \ldots,\mathbf{z}_{\delta}(t)^T]$.

Equations~(\ref{eq:decoupledEqs}) correspond to a set of $\mu$ decoupled systems. Each of them can be further decoupled by considering the state transformation $\xi_i(t)=(\mathrm{V}_i^T \otimes \mathrm{I}_2)\mathbf{z}_i(t)$, where $\mathrm{V}_i$ is the orthogonal matrix diagonalizing $\mathrm{R}_i \tilde{\mathrm{L}}_i$, i.e., $\mathrm{R}_i \tilde{\mathrm{L}}_i= \mathrm{V}_i \tilde{\mathrm{\Lambda}}_i \mathrm{V}_i^T$ where $\tilde{\mathrm{\Lambda}}_i$ is the diagonal matrix containing the eigenvalues of the $\mathrm{R}_i \tilde{\mathrm{L}}_i$. In addition, we have to take into account that for each block $i$ with $i=1,\ldots,\mu$ the  matrices $\mathrm{R}_i \tilde{\mathrm{L}}_i$ and $\tilde{\mathrm{L}}_i$ have the same eigenvalues. To show this, consider the block decomposition of $\mathrm{R}\tilde{\mathrm{L}}$ as in (\ref{eq:RLdecomposition}). From this block decomposition, one derives that the eigenvalues of $\mathrm{R}\tilde{\mathrm{L}}$ are those of the blocks appearing in its main diagonal. For each of the first $\mu$ blocks, we note that the two matrices, $\mathrm{R}_i \tilde{\mathrm{L}}_i$ and $\tilde{\mathrm{L}}_i$, have the same eigenvalues. In fact, from Lemma~\ref{lemma:sameEigenv}, we have that $\lambda_j(\mathrm{R}_i \tilde{\mathrm{L}}_i)= \lambda_j(\tilde{\mathrm{L}}_i \mathrm{R}_i)$, and, since $\mathrm{R}_i=\mathrm{I}_{n_i}-\frac{1}{n_i}\mathbf{1}_{n_i}\mathbf{1}_{n_i}^T$, we have that $\tilde{\mathrm{L}}_i \mathrm{R}_i= \tilde{\mathrm{L}}_i - \frac{1}{n_i}(\tilde{\mathrm{L}}_i \mathbf{1}_{n_i})\mathbf{1}_{n_i}^T= \tilde{\mathrm{L}}_i$. Hence, $\lambda_j(\mathrm{R}_i \tilde{\mathrm{L}}_i)= \lambda_j(\tilde{\mathrm{L}}_i \mathrm{R}_i)=\lambda_j(\tilde{\mathrm{L}}_i)$. Finally, as $\lambda_1(\tilde{\mathrm{L}}_i)=0$ and the corresponding eigenvector is parallel to $\mathbf{1}_{n_i}$, then $\mathrm{V}_i^T \mathrm{R}_i\mathrm{V}_i=\mathrm{I}_{n_i}-\frac{1}{n_i}\mathbf{e}^{n_i}_1\mathbf{e}^{n_i,T}_1$, where $\mathbf{e}^{n_i}_1$ ($i=1,\ldots,n_i$) is the standard basis of $\mathbb{R}^{n_i}$. Altogether, these considerations yield

\begin{equation}
\label{eq:decoupledEqs3}
\dot{\mathbf{\xi}}_i(t)= \mathrm{A}\Delta_i \mathbf{\xi}_i(t) - \lambda_j(\tilde{\mathrm{L}}_i) \mathrm{B}\mathrm{K} \mathbf{\xi}_i(t) \\
\end{equation}

\noindent with $i=1,\ldots,\mu$, $j=1,\ldots,n_i$, and $\Delta_i=0$ if $i=1$, and
$\Delta_i=1$ otherwise. Hence, for $j=1$, since $\lambda_1(\tilde{\mathrm{L}}_i)=0$, we obtain the mode along the multi-consensus manifold, while for $j=2,\ldots,n_i$ the modes transverse to it. So, stability of the transverse modes in Eq.~(\ref{eq:decoupledEqs3}) is studied by considering the characteristic equation $\det(s\mathrm{I}_2-\mathrm{A}+\lambda_j(\tilde{\mathrm{L}}_i)\mathrm{B} \mathrm{K})$ and correspondingly the polynomials $f_j(s)=s^2-[b-k_2 \lambda_j(\tilde{\mathrm{L}}_i) ]s-[a-k_1\lambda_j(\tilde{\mathrm{L}}_i)]$, for $i=1,\ldots,\mu$ and $j=2,\ldots,n_i$.

To have stable dynamics the roots of $f_j(s)=0$ with $j=2,\ldots,n_i$ must be in the left half-plane, thus applying the Routh-Hurwitz criterion we obtain that the control gains have to satisfy that $k_1>\frac{a}{\lambda_2(\tilde{\mathrm{L}}_i)}$ and $k_2>\frac{b}{\lambda_2(\tilde{\mathrm{L}}_i)}$.

A further condition derives from the inspection of Eq.~(\ref{eq:decoupledEqsRM}). This system can be viewed as a forced system, where the inputs are $\mathbf{Z}_{h}(t)$ with $h=1,\ldots,\mu$. If the system~(\ref{eq:decoupledEqsRM}) is stable and so are those of Eqs.~(\ref{eq:decoupledEqs3}), then the inputs of~(\ref{eq:decoupledEqsRM}) converge to zero and $\mathbf{Z}_{\mu+1}(t)\rightarrow 0$.

The stability of the transverse modes of system~(\ref{eq:decoupledEqsRM}) is studied similarly to (\ref{eq:decoupledEqs3}), yielding the conclusion that $k_1>\frac{a}{\lambda_2(\mathrm{R_{\delta}}\mathrm{M}_{\delta})}$ and $k_2>\frac{b}{\lambda_2(\mathrm{R_{\delta}}\mathrm{M}_{\delta})}$.

The eigenvalues of $\mathrm{R_{\delta}}\mathrm{M}_{\delta}$ are also related to those of $\tilde{\mathrm{L}}$. Consider, in fact, Lemma~\ref{lem:eigproduct} with $\pi=\pi^*$, then it immediately follows that the non-zero eigenvalues of $\mathrm{R}\mathrm{M}_{\delta}$ are $\{\lambda_j(\tilde{\mathrm{L}})\}
\setminus \{\lambda_j(\tilde{\mathrm{L}}_{\pi^*})\}\setminus \{\lambda_j(\tilde{\mathrm{L}}_1)\}\setminus \ldots \setminus \{\lambda_j(\tilde{\mathrm{L}}_\mu)\}$.

As for multi-consensus the stability of the transverse modes of both (\ref{eq:decoupledEqs}) and (\ref{eq:decoupledEqsRM}) is required and $\tilde{\mathrm{L}}$ and $\mathrm{L}$ are similar, the thesis follows. $\square$

\begin{remark}
\label{rem:partialconsensus}
In Theorem~\ref{th:mainforthesecondorder} the stability of the multi-consensus state is obtained if $k_1>\frac{a}{\lambda_2(\tilde{\mathrm{L}}_i)}$ and $k_2>\frac{b}{\lambda_2(\tilde{\mathrm{L}}_i)}$ with $i=1,\ldots,\mu$ and if $k_1>\frac{a}{\lambda_2(\mathrm{R_{\delta}}\mathrm{M}_{\delta})}$ and $k_2>\frac{b}{\lambda_2(\mathrm{R_{\delta}}\mathrm{M}_{\delta})}$. The condition is, therefore, checked for each of the clusters $\mathcal{H}_i$, $i=1,\ldots,\mu$, associated with the reachable sets and for the clusters deriving from the subdivision in cells $\mathcal{C}_i$, $i=1,\ldots,h$ of the common part $\mathcal{C}$. Failure of the stability condition in a single cluster clearly leads to the loss of multi-consensus. However, as the clusters $\mathcal{H}_i$ with $i=1,\ldots,\mu$, are independent each other and from the cells $\mathcal{C}_i$, a regime of \emph{partial} consensus may be observed with some clusters converging to a common value, while the others not.
\end{remark}

\subsection{Numerical examples}

\emph{Example 3}. Let us consider a multi-agent system with second-order dynamics as in Eqs.~(\ref{eq:IIorder}) with $a=1$ and $b=0.8$ and a network of interaction as in Example 1 and Fig.~\ref{fig:graph1}(a). The target multi-consensus is the same as in Example 1 and so is the topology of the controlled multi-agent system obtained with the same steps of the case of single integrator dynamics. We focus here on the stability condition which differs from the scenario previously studied.

\begin{figure}
\begin{center}
\subfigure[]{\includegraphics[width=0.24\textwidth]{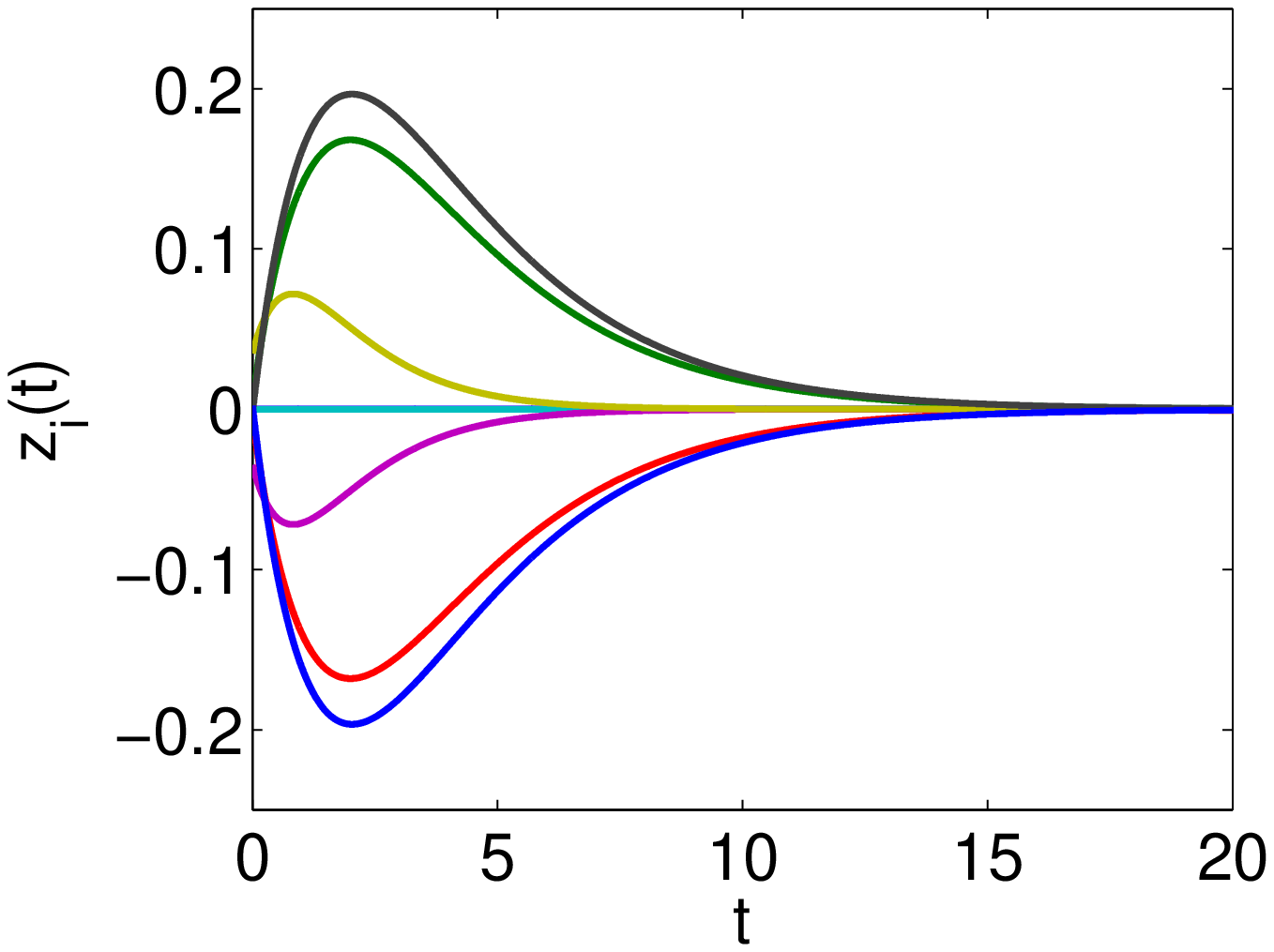}}
\subfigure[]{\includegraphics[width=0.24\textwidth]{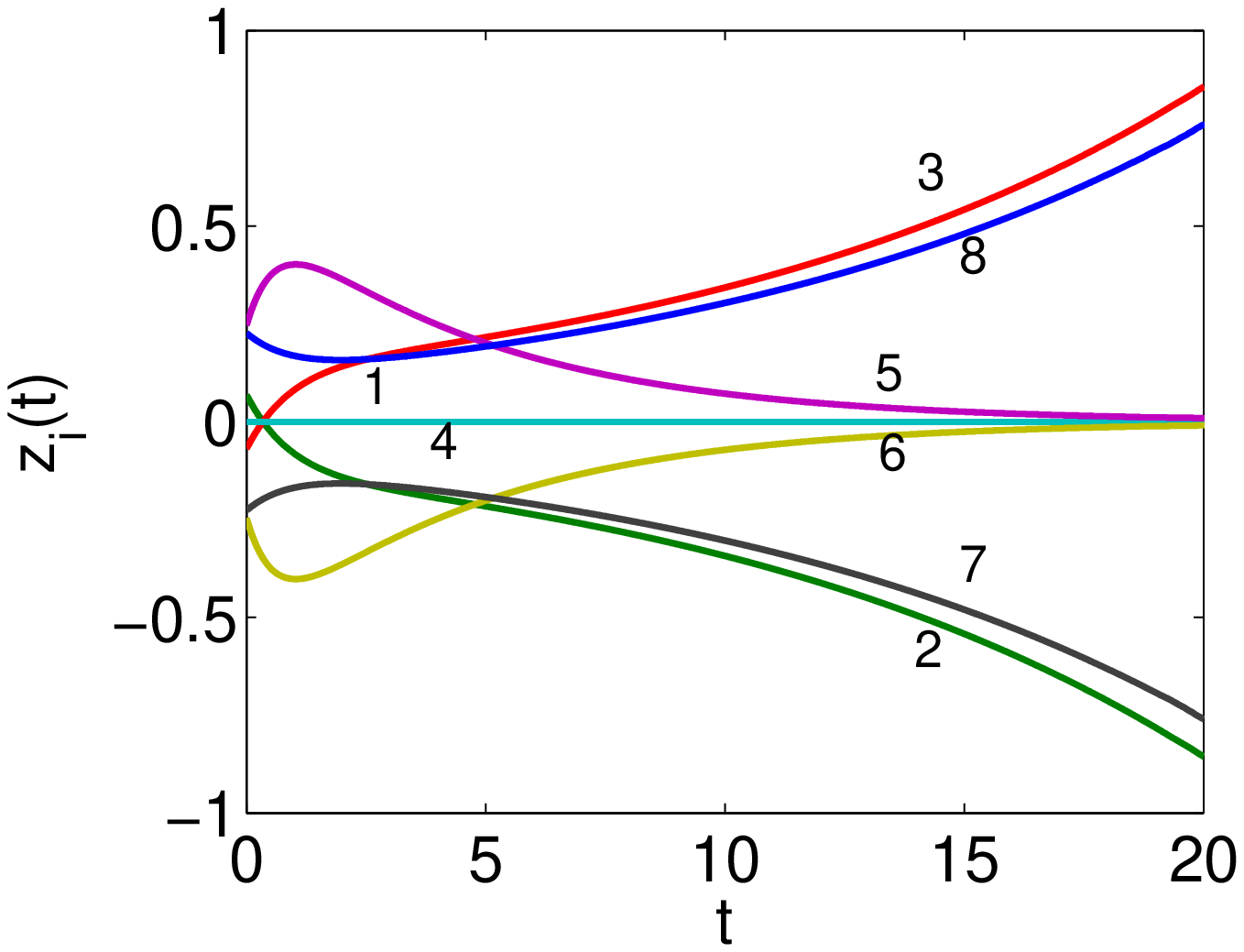}}
\caption{\label{fig:multicluster} Time evolution of the variables $\mathbf{z}_i(t)$ for the second-order consensus protocol as in Eq.~(\ref{eq:IIorder}): (a) multi-cluster consensus with $k_1=0.62$ and $k_2=0.98$; (b) only the cluster \{5,6\} is stable for $k_1=0.45$ and $k_2=0.98$. The network of interaction of the agents of the system is as in Example 1 and Fig.~\ref{fig:graph1}(a).}
\end{center}
\end{figure}

To find $\gamma$ of Theorem~\ref{th:mainforthesecondorder}, we calculate the eigenvalues of $\mathrm{L}+\mathrm{L}^u$ (or equivalently those of $\tilde{\mathrm{L}}$) and of the Laplacian of the quotient graph. We obtain: $\{\lambda_i(\mathrm{L}+\mathrm{L}^u)\}=\{0,0,0,2,2,2,3,3\}$ and $\{\lambda_i((\mathrm{L}+\mathrm{L}^u)_{\pi^*})\}=\{0,0,0,2,3\}$, so that $\gamma=2$. The stability condition is therefore $k_1>\frac{a}{2}=0.5$ and $k_2>\frac{b}{2}=0.4$. Selecting, for instance,
$k_1=0.62$ and $k_2=0.98$ results in a stable multi-cluster as shown in Fig.~\ref{fig:multicluster}(a).

As in the proof of Theorem~\ref{th:mainforthesecondorder}, the stability of each cluster may also be studied. To do this, we have to consider the eigenvalues of the corresponding block in the Laplacian matrix that, for this example, is given by (\ref{eq:laplacianEx1}). For each block, stability of the cluster $i$ requires that $k_1>\frac{a}{\lambda_2(\tilde{\mathrm{L}}_i)}$ and $k_2>\frac{a}{\lambda_2(\tilde{\mathrm{L}}_i)}$ if $i=1,\ldots,3$ or $k_1>\frac{a}{\lambda_2(\mathrm{R_{\delta}}\mathrm{M}_{\delta})}$ and $k_2>\frac{a}{\lambda_2(\mathrm{R_{\delta}}\mathrm{M}_{\delta})}$ if the block is the one associated to the common part. The eigenvalues of each block are $\{\lambda_i(\tilde{\mathrm{L}}_1)\}=\{0\}$, $\{\lambda_i(\tilde{\mathrm{L}}_2)\}=\{\lambda_i(\tilde{\mathrm{L}}_3)\}=\{0, 2\}$, and $\{ \lambda_i(\mathrm{R_{\delta}}\mathrm{M}_{\delta})\}=\{0,0,3\}$. From this, we retrieve the previously found stability condition, i.e., $k_1>\frac{a}{2}=0.5$ and $k_2>\frac{b}{2}=0.4$.

Suppose now to select $0.3333=\frac{a}{3}<k_1<\frac{a}{2}=0.5$ and $k_2>\frac{b}{2}=0.4$, then the multi-consensus is not stable, but the cluster formed by nodes $\{5,6\}$ is stable (Remark~\ref{rem:partialconsensus}). An example of this latter case is shown in Fig.~\ref{fig:multicluster}(b), obtained for $k_1=0.45$ and $k_2=0.98$.

\begin{figure}
\begin{center}
\subfigure[]{\includegraphics[width=0.24\textwidth]{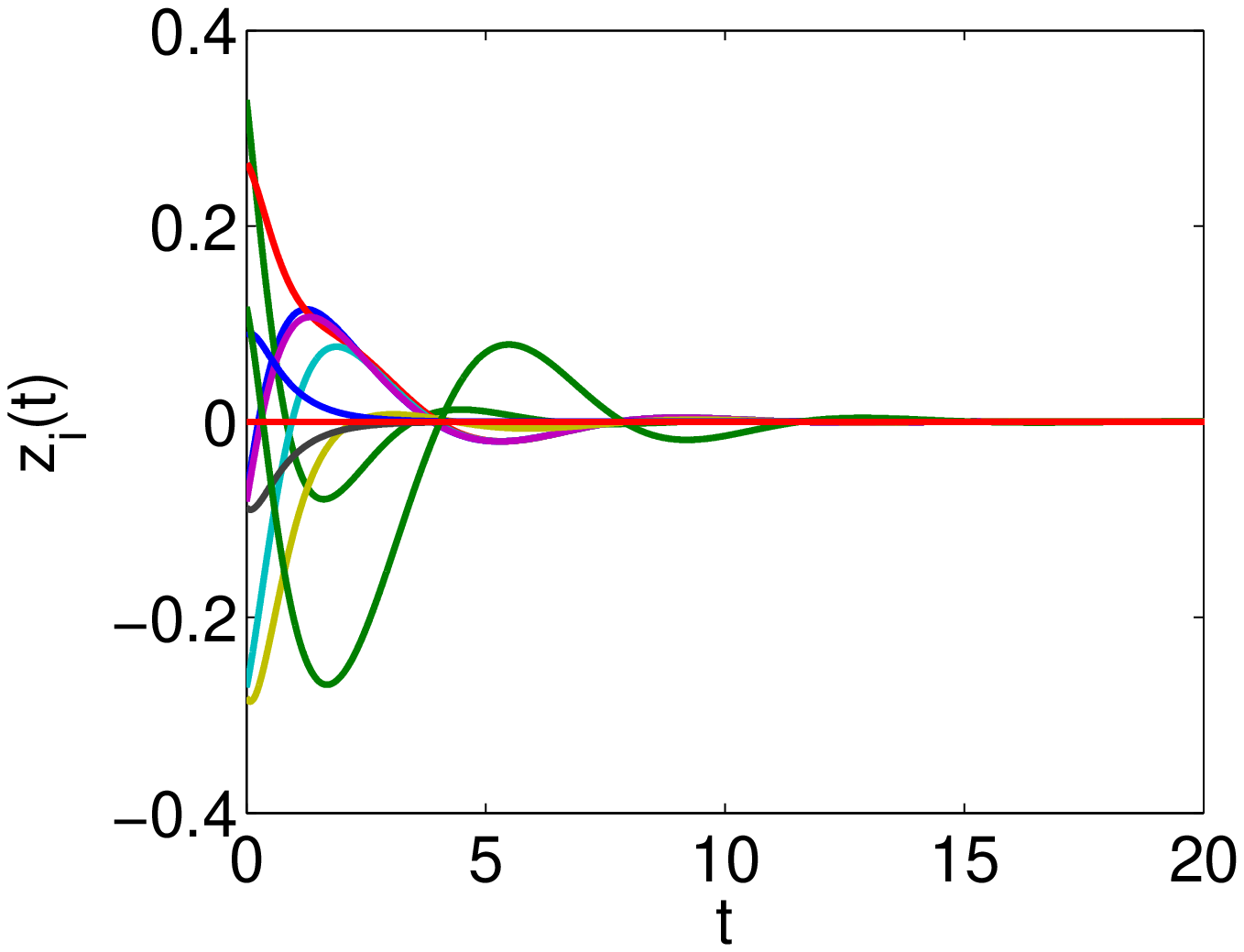}}
\subfigure[]{\includegraphics[width=0.24\textwidth]{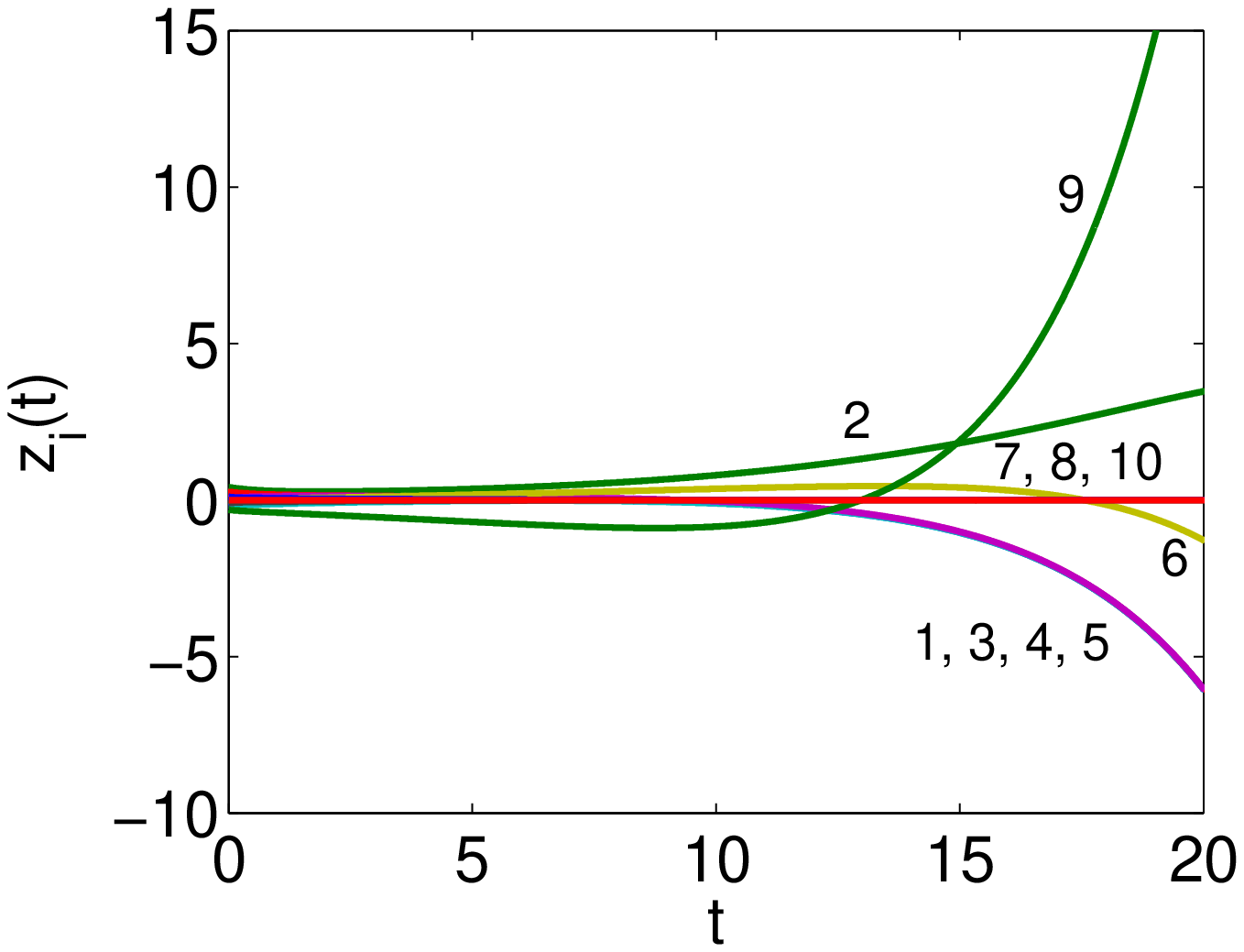}}
\caption{\label{fig:multiclusterII} Time evolution of the variables $\mathbf{z}_i(t)$ for second-order consensus protocol as in Eq.~(\ref{eq:IIorder}): (a) multi-cluster consensus with $k_1=2$ and $k_2=1.8$; (b) only the cluster \{7,8\} is stable for $k_1=0.4$ and $k_2=1.8$. The network of interaction of the agents of the system is as in Example 2 and Fig.~\ref{fig:graph2}(a)}
\end{center}
\end{figure}

\emph{Example 4}. We consider now the multi-agent system with second-order dynamics as in Eqs.~(\ref{eq:IIorder}) with $a=1$ and $b=0.8$ and a network of interaction as in Example 2 and Fig.~\ref{fig:graph2}(a). The target multi-consensus is as in Example 2, thus resulting in the same network for the controlled multi-agent system.
In this case, $\gamma=1$, as $\{\lambda_i(\mathrm{L}+\mathrm{L}^u)\}=\{0,0,1,1,2,2,2,3,3,4\}$ and $\{\lambda_i((\mathrm{L}+\mathrm{L}^u)_{\pi^*})\}=\{0,0,3\}$. It follows that if $k_1>a=1$ and $k_2>b=0.8$ the system reaches multi-consensus, as shown in Fig.~\ref{fig:multiclusterII}(a) for $k_1=2$ and $k_2=1.8$.

The eigenvalues for the different blocks appearing in the Laplacian (given by Eq.~(\ref{eq:laplacianEx2})) associated to this network are $\{\lambda_i(\tilde{\mathrm{L}}_1)\}=\{0,1,1,2,2,2,3\}$, $\{\lambda_i(\tilde{\mathrm{L}}_2)=\{0\}\}$, and $\{\lambda_i(\mathrm{R_{\delta}}\mathrm{M}_{\delta})\}=\{0,4\}$. Taking into account that the smallest non-zero eigenvalue of the block $\mathrm{R_{\delta}}\mathrm{M}_{\delta}$ is $\lambda_2=4$ and choosing the control gains as $0.25=\frac{a}{4}<k_1<a=1$ and $k_2>\frac{b}{4}=0.4$ only the cluster formed by nodes $\{7,8\}$ is stable, as shown in Fig.~\ref{fig:multiclusterII}(b) for $k_1=0.4$ and $k_2=1.8$.

\section{Extension to signed Laplacians}
\label{sec:signedLaplacians}

In this Section we consider the case where the controllers may either add new links or remove some existing ones. In this scenario, the matrix $\mathrm{L}^u$ is no more a Laplacian matrix as in Lemma~\ref{lem:lemma1}, but it is a signed Laplacian \cite{boyd2006convex}, namely its generic element $\mathrm{L}_{ij}^u$ can be: $\mathrm{L}_{ij}^u=-1$ if a new link is added from $j$ to $i$; $\mathrm{L}_{ij}^u=1$ if an existing link from $j$ to $i$ is removed; $\mathrm{L}_{ij}^u=0$ if no change to the connection from $j$ to $i$ is made by the controllers. The diagonal elements are such that the Laplacian is still a zero-row sum matrix, i.e., $\mathrm{L}_{ii}^u=-\sum\limits_j\mathrm{L}_{ij}^u$. Also note that, since we allow to remove a link only if it exists in the original topology, the matrix $\mathrm{L}+\mathrm{L}^u$ is still a Laplacian in the classical sense. We now discuss a generalization of Lemma~\ref{lem:lemma1} to this scenario.

\begin{lemma}
\label{lem:lemma1signed}
Given a network with Laplacian matrix $\mathrm{L}$ and a partition $\pi^Q$, there exists a \emph{signed} Laplacian matrix $\mathrm{L}^u$ such that $\pi^Q$ is an EEP for the network with Laplacian matrix $\mathrm{L}+\mathrm{L}^u$.
\end{lemma}

\emph{Proof:}
The proof follows the same steps of Lemma~\ref{lem:lemma1} with the main difference regarding the associated optimization problem. The binary variables $y_h$ with $h=(i-1)N+j$ of the problem now have the following meaning:
if $a_{ij}=0$, $y_h=1$ indicates that a link has to be added; if $a_{ij}=1$, $y_h=1$ indicates that the existing link has to be removed; $y_h=0$ indicates no change. Let us define $\delta_h$ as: $\delta_h=-1$ if $a_{ij}=1$, and $\delta_h=1$ otherwise, and indicate the columns of the matrix $\mathrm{M}$ appearing in Eq.~(\ref{eq:MxB}) as $\mathrm{M}_1,\ldots,\mathrm{M}_h,\ldots,\mathrm{M}_{N^2}$. Let us also consider a new matrix, $\mathrm{\bar{M}}$ 
defined as $\bar{\mathrm{M}}=\delta_1\mathrm{M}_1,\ldots,\delta_h\mathrm{M}_h,\ldots,\delta_{N^2}\mathrm{M}_{N^2}$, then (\ref{eq:permute2}) is rewritten as:
\begin{equation}
\mathrm{\bar{M}}y=B
\end{equation}

Correspondingly, the optimization problem is defined as:
\begin{equation}
\label{eq:LyapEqVecV2minpSigned}
\min f^T y, \mathrm{~subject~to~}\mathrm{\bar{M}} y=\mathrm{B}
\end{equation}
Once obtained the solution $y$ of the optimization problem the generic element of the signed Laplacian $\mathrm{L}^u$ is given by: $\mathrm{L}_{ij}^u=-\delta_h y_h$ with $h=(i-1)N+j$.
$\square$

Lemma~\ref{lem:lemma1signed} does not guarantee that the connectivity of the network is preserved as removing links from the original structure can result in a new network with some cluster isolated from the rest of the network. The following Lemma incorporates a further constraint in the optimization problem that guarantees that the network remains weakly connected.

\begin{lemma}
\label{lem:lemma1signedconnected}
Given a weakly connected network with Laplacian matrix $\mathrm{L}$ and a partition $\pi^Q$, there exists a \emph{signed} Laplacian matrix $\mathrm{L}^u$ such that the network with Laplacian matrix $\mathrm{L}+\mathrm{L}^u$ is weakly connected and $\pi^Q$ is an EEP for it.
\end{lemma}

\emph{Proof:} Also in this case, the proof follows the same steps of Lemma~\ref{lem:lemma1} and Lemma~\ref{lem:lemma1signed}, so we discuss only the new constraints that need to be incorporated in the optimization problem.

Two generic cells $C_m$ and $C_k$ of the partition $\pi^Q$ with $m,k=1,\ldots,Q, m\neq k$ are connected if

\begin{equation}
\label{eq:doppiasomma}
\sum\limits_{i\in C_m}\sum\limits_{j\in C_k}(-L_{ij}-L_{ij}^u)>0
\end{equation}

Taking into account that $\mathrm{L}_{ij}^u=-\delta_h y_h$, (\ref{eq:doppiasomma}) can be rewritten as

\begin{equation}
\label{eq:furtherlink}
\sum\limits_{h=(i-1)N+j|i\in C_m,j\in C_k}\delta_h y_h>\sum\limits_{i\in C_m}\sum\limits_{j\in C_k}L_{ij}
\end{equation}

\noindent with $m,k=1,\ldots,Q, m\neq k$. The optimization problem incorporating the constraints (\ref{eq:furtherlink}) guarantees the weak connectivity of the network with Laplacian matrix $\mathrm{L}+\mathrm{L}^u$. It reads:

\begin{equation}
\label{eq:LyapEqVecV2minpSignedConnected}
\min f^T y, \mathrm{~subject~to~}\mathrm{\bar{M}} y=\mathrm{B} \mathrm{~and~}(47)
\end{equation}
$\square$

The design of the controllers for agents with single integrator or with second-order dynamics, in the case where links can be either added or removed, eventually maintaining the original weak connectedness, is performed by using Theorem~\ref{th:th1} or Theorem~\ref{th:mainforthesecondorder}, replacing Lemma~\ref{lem:lemma1} with Lemma~\ref{lem:lemma1signed} or with Lemma~\ref{lem:lemma1signedconnected} in the step to find $\mathrm{L}^u$.

\subsection{Numerical examples}

\begin{figure}
\subfigure[]{\includegraphics[width=0.24\textwidth]{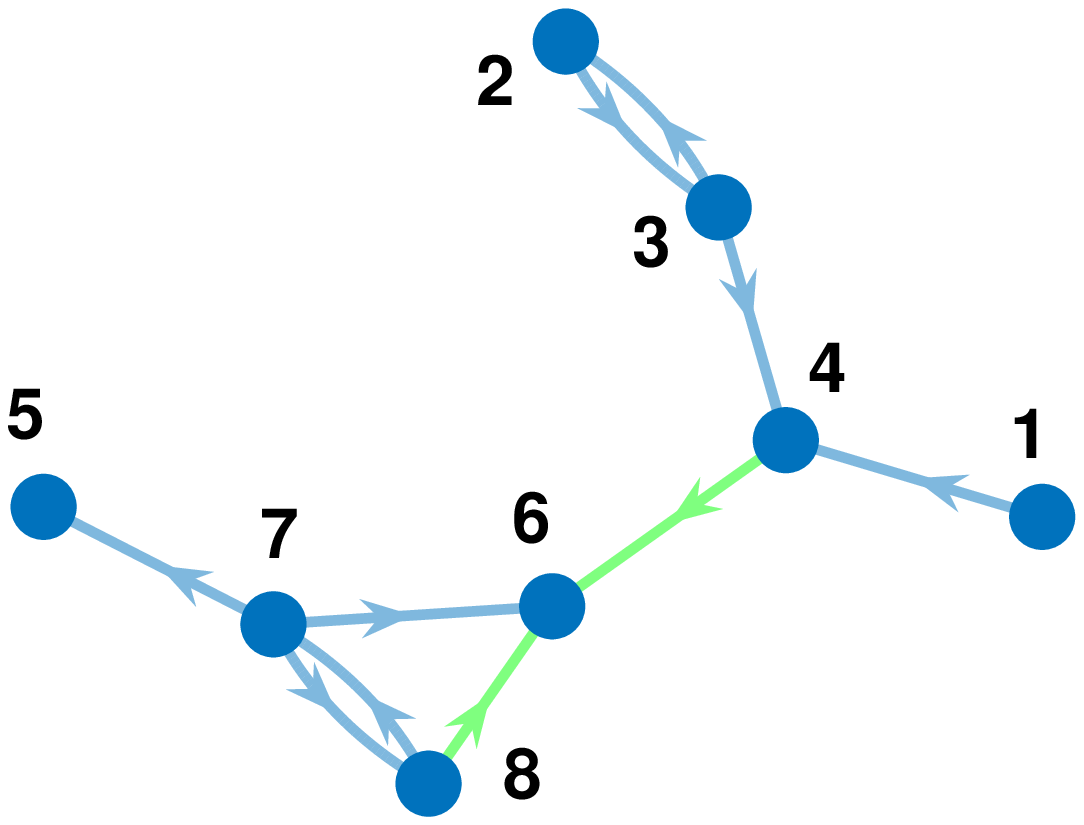}}
\subfigure[]{\includegraphics[width=0.24\textwidth]{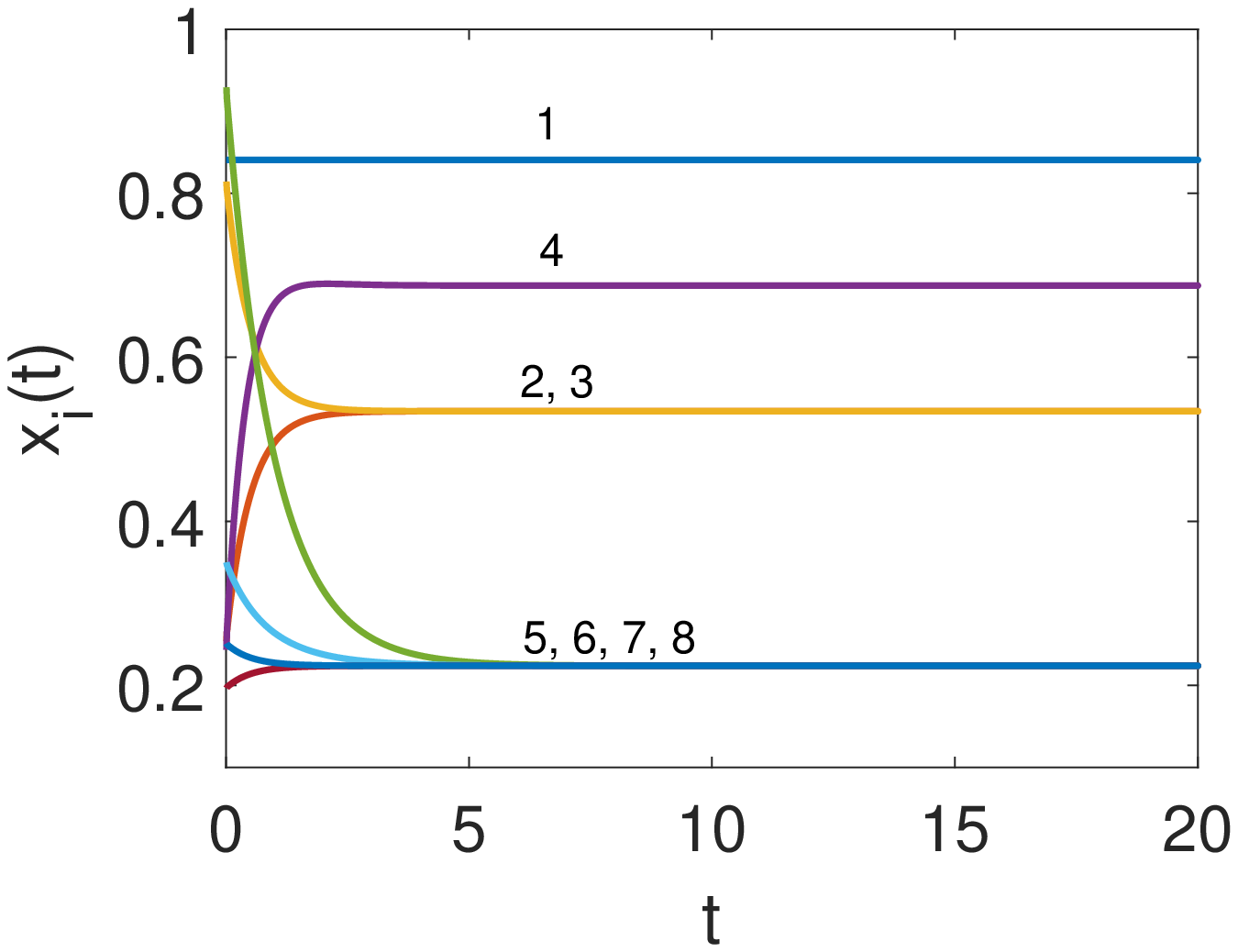}}
\caption{\label{fig:Esempio1ConSegno} Multi-consensus of single integrators via Lemma~\ref{lem:lemma1signed}. (a) Digraph with $N=8$ modeling the interactions among the agents. In blue the links of the original graph are shown, in green those removed to reach the desired multi-consensus as in Example~1. (b) Time evolution of variables $\mathbf{x}_i(t)$.}
\end{figure}

\emph{Example 5}. Let us consider again the multi-agent system and multi-consensus problem as in Example 1 and apply the method in Lemma~\ref{lem:lemma1signed} to find the controllers. Fig.~\ref{fig:Esempio1ConSegno} shows the results. We notice that the digraph obtained (Fig.~\ref{fig:Esempio1ConSegno}(a)) is no more weakly connected as the link (4,6) has been removed. The time evolution of the variables $\mathbf{x}_i(t)$ confirms that the system reaches the desired multi-consensus.

\begin{figure}
\subfigure[]{\includegraphics[width=0.24\textwidth]{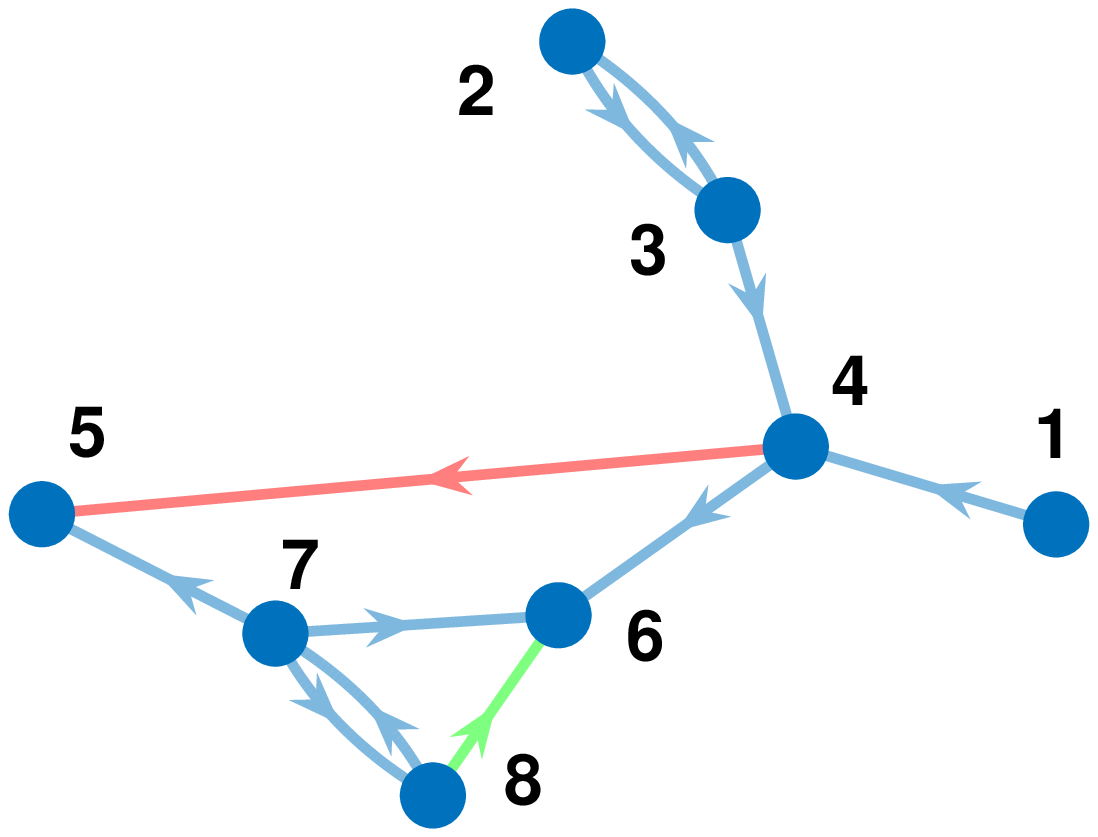}}
\subfigure[]{\includegraphics[width=0.24\textwidth]{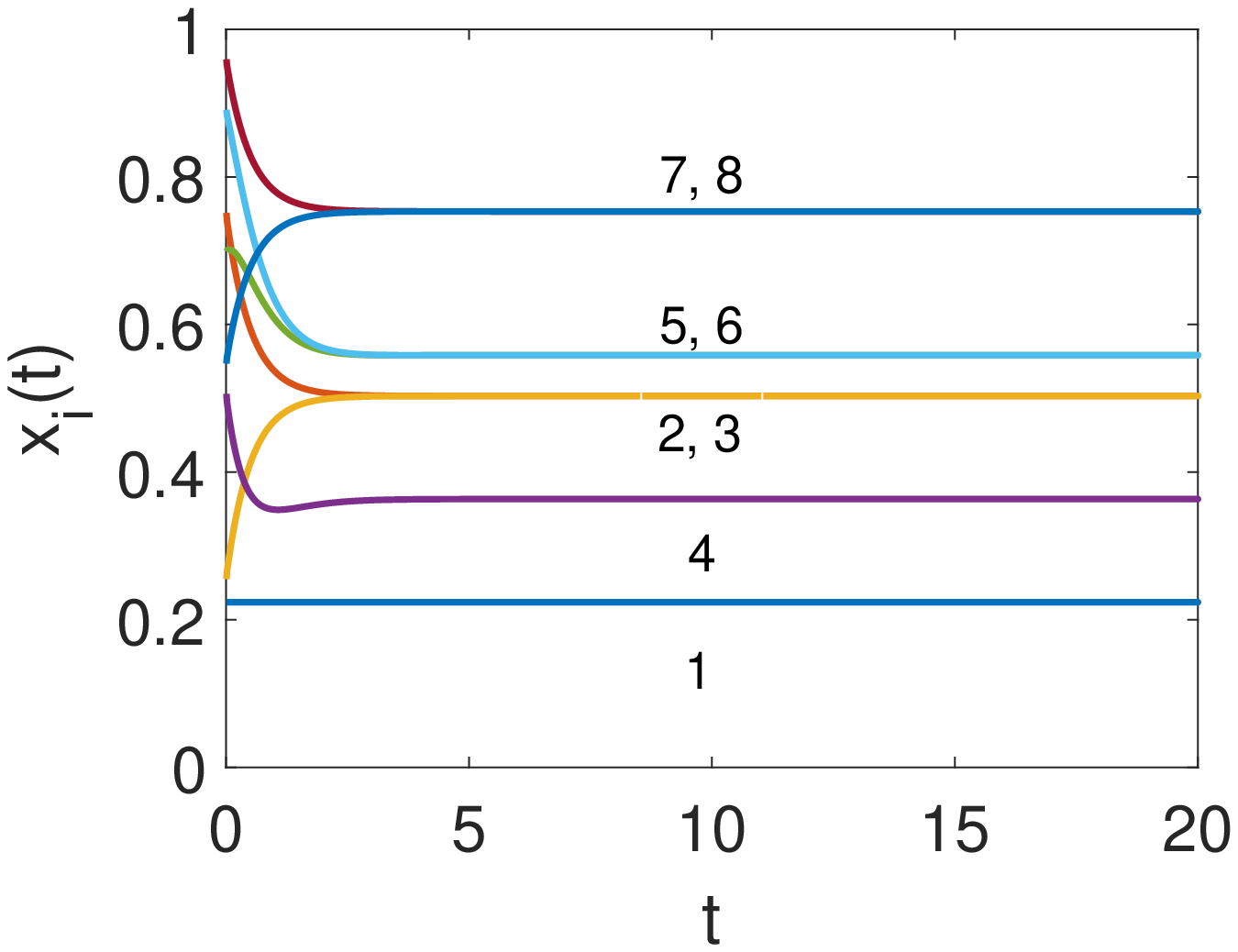}}
\caption{\label{fig:Esempio1ConSegnoConConn} Multi-consensus of single integrators via Lemma~\ref{lem:lemma1signedconnected}. (a) Digraph with $N=8$ modeling the interactions among the agents. In blue the links of the original graph are shown, in red (green) those added (removed) to reach the desired multi-consensus as in Example~1. (b) Time evolution of variables $\mathbf{x}_i(t)$.}
\end{figure}

\emph{Example 6}. In this case, we apply the method in Lemma~\ref{lem:lemma1signedconnected} to the problem in Example~1, such that the controllers may add or remove links, while preserving weak connectedness. The result is illustrated in Fig.~\ref{fig:Esempio1ConSegnoConConn}. In the digraph obtained (Fig.~\ref{fig:Esempio1ConSegnoConConn}(a)) a link has been removed and another added. The system reaches the desired multi-consensus as shown in Fig.~\ref{fig:Esempio1ConSegnoConConn}(b).

\section{Conclusions}
\label{sec:conclusions}

In this paper, we have studied the problem of multi-consensus control. Given a multi-agent system with an underlying digraph of interaction among the units and a desired multi-consensus, we have shown that distributed control may be applied to drive the system towards the target regime. The design of the controllers consists of two steps. The first step is based on the fulfillment of a topological condition, i.e., the existence of an external equitable partition, which is equivalently reformulated in terms of an algebraic condition on the network Laplacian. This step is addressed by formulating three integer linear programming problems that arise considering different requirements on the network of interactions among agents. In the first case, only the addition of links is considered and a constructive algorithm solving the integer linear programming problem has been also discussed. In the remaining cases, addition and removal of links in the original structure of interactions are considered. As removing links may result in a new network loosing the original property of weak connectedness, we have proposed two methods, accounting for the cases where maintaining the connectedness is important or not. From a mathematical point of view, the important difference in the three scenarios is that the matrix $\mathrm{L}^u$ is a Laplacian in the classical sense when links are exclusively added, while it is a signed Laplacian in the remaining cases. In all the three cases, the minimum set of links that need to be changed in the original topology is obtained.

The second step concerns the stability of the multi-consensus state. In the case of single integrator dynamics, stability is guaranteed without further requirements as a consequence of the positive semidefiniteness of the Laplacian matrix, while, in the case of second-order dynamics, stability requires a condition on the gains used in the communication protocol. This condition has been analytically derived in this work and numerical examples reported to illustrate it.

As multi-consensus offers more flexibility than consensus in allowing agents to split into groups reaching different consensus values, applications where multi-agent systems are required to perform multiple tasks in parallel or to perform simultaneous measurements of a variable in different areas, may benefit of strategies for the control of this state. An example of such applications can be intentional islanding in power grids, which describes a condition where a portion of the network is isolated from the remainder of the system and it is important to guarantee the normal operation (usually identified with the synchronous state) of the isolated portion of network.

Noticeably, the proposed approach relies on the configuration of a proper communication protocol among agents, similarly to what is done for consensus, so that it is possible to envisage a scenario where the multi-agent system is reconfigured to reach consensus or one or more multi-consensus states only by intervening on its communication protocol.

\section*{Acknowledgment}

This work was supported by the Italian Ministry for Research and Education (MIUR) through Research Program PRIN 2017 under Grant 2017CWMF93, project VECTORS. The Authors would like to thank all the anonymous Reviewers for their comments and suggestions. We also acknowledge that the core idea of the constructive algorithm for the solution of the optimization problem (\ref{eq:LyapEqVecV2minp}) was suggested by one of the Reviewers.

\bibliographystyle{IEEEtran}


~\\

\end{document}